\documentclass[twocolumn]{aastex7}

\usepackage{mhchem}
\usepackage{bm}
\usepackage{subcaption}
\usepackage{threeparttable}
\usepackage{tabularx}
\usepackage{booktabs}
\usepackage{enumitem}
\usepackage{placeins}
\begin{document}

\title{CORINOS V: Radiative transfer effects in protostellar ice observations}

\author[0000-0002-1353-2562]{Will E. Thompson}
\affiliation{University of California, Berkeley, Berkeley, CA, 94720, USA} 
\email{willet@berkeley.edu}

\author[0000-0002-8716-0482]{Jennifer B. Bergner}
\affiliation{University of California, Berkeley, Berkeley, CA, 94720, USA} 
\email{}

\author[0000-0001-5175-1777]{Neal J. Evans II}
\affiliation{Department of Astronomy, The University of Texas at Austin, Austin, TX 78712, USA} 
\email{}

\author[0000-0001-8227-2816]{Yao-Lun Yang}
\affiliation{RIKEN Cluster for Pioneering Research, Wako-shi, Saitama, 351-0198, Japan} 
\email{}

\author[0009-0004-8301-7947]{Vincent Kreft}
\affiliation{University of California, Berkeley, Berkeley, CA, 94720, USA} 
\email{}

\author[0009-0001-0753-2937]{Lenore Anderson}
\affiliation{University of California, Berkeley, Berkeley, CA, 94720, USA} 
\email{} 

\author[0000-0001-7552-1562]{Klaus M. Pontoppidan}
\affiliation{Jet Propulsion Laboratory, California Institute of Technology, 4800 Oak Grove Drive, Pasadena, CA 91109, USA} 
\email{} 

\author[0000-0003-2076-8001]{L.Ilsedore Cleeves}
\affiliation{University of Virginia, Department of Astronomy, Charlottesville, VA 22904, USA}
\affiliation{University of Virginia, Department of Chemistry, Charlottesville, VA 22904, USA}
\email{}

\author[0000-0001-7591-1907]{Ewine F. van Dishoeck}
\affiliation{Leiden Observatory, Leiden University, P. O. Box 9513, 2300 RA Leiden, The Netherlands}
\affiliation{Max Planck Institut für Extraterrestrische Physik (MPE), Giessen-
bachstrasse 1, 85748 Garching, Germany}
\email{}

\author[0000-0002-0477-6047]{Rachel E. Gross}
\affiliation{University of Virginia, Department of Chemistry, Charlottesville, VA 22904, USA}
\email{}

\author[0000-0003-3119-2087]{Jeong-Eun Lee}
\affiliation{Department of Physics and Astronomy, Seoul National University, 1 Gwanak-ro, Gwanak-gu, Seoul 08826, Korea}
\email{}

\author[0000-0003-1878-327X]{Melissa K. McClure}
\affiliation{Leiden Observatory, Leiden University, PO Box 9513, NL–2300 RA Leiden, The Netherlands}
\email{}

\author[0000-0002-3297-4497]{Nami Sakai}
\affiliation{RIKEN Cluster for Pioneering Research, Wako-shi, Saitama, 351-0198, Japan}
\email{}

\author[0000-0002-7433-1035]{Katerina Slavicinska}
\affiliation{Laboratory for Astrophysics, Leiden Observatory, P.O. Box 9513, 2300 RA Leiden, NL}
\email{}

\begin{abstract}
Recent observations of protostars with the James Webb Space Telescope have revealed unprecedented chemical complexity from their ice absorption features. However, these spectra are likely influenced by radiative transfer effects, and there is little understanding of how this impacts our ability to identify, quantify, and interpret the observed ice features. We have developed a new modeling framework to investigate the radiative transfer through icy protostellar envelopes, and apply this to the IRAS 15398-3359 protostar observed by the JWST CORINOS program. The modeled H$_2$O and CO column densities are similar to previous empirical studies, but we require a high \ce{CO2}/\ce{H2O} ratio of 76\% to match the optical depth of the 15 $\mu$m band.  We use our modeled continuum to calculate a 6--10 $\mu$m optical depth spectrum, and see considerable differences compared to a simple polynomial continuum model, underscoring the challenges with quantifying trace ice species in this range.  For this source, we find that the observed absorption predominantly originates along the viewing line of sight between 1000 -- 2000 au, peaking at the transition from the outflow cavity to the envelope; the spectra are largely insensitive to absorption from ices in the outer envelope, which extends out to 20,000 au.  
Lastly, we show that depending on how the line of sight intersects the cavity, the apparent CO$_2$/H$_2$O and CO/H$_2$O column density ratios can be underestimated compared to the underlying ice abundance ratios. Together this provides important context for interpreting the ice constraints derived from JWST observations of protostars.

\end{abstract}


\section{Introduction} \label{sec:intro}

Star and planet formation commences with the collapse of a dense core within a molecular cloud, leading to a central embedded protostar within an envelope of gas and dust \citep{Shu1987, McKee2007, Andrews2018}. The infall of envelope material eventually leads to the formation of an accretion disk, within which planet formation takes place.  Therefore, the compositions of planets and planetesimals will be shaped by the chemical and physical conditions of the protostellar envelope \citep{Visser2009, Oberg2011, Henning2013}. 

Radiation from the protostar that travels through the surrounding envelope is absorbed and scattered by dust grains \citep{Whitney2003, Yang2017}.  The absorption and scattering efficiencies of a dust grain population are influenced by the grain physical properties, including the size distribution, grain shapes, and chemical composition.  Constraints on these dust properties in the diffuse interstellar medium, molecular clouds, and young stellar objects can be obtained from spectroscopic and photometric observations \citep{Ossenkopf1994, Dartois2024}. This informs our knowledge of the optical properties of dust grains, and in turn the thermal and physical structure and chemical evolution of protostars  \citep{Evans1999, Weingartner2001, Draine2003, Boogert2015}.

Infrared observations towards young stellar objects have shown that the dust grains, themselves composed largely of silicate and carbonaceous refractory material, are also commonly covered by ice mantles \citep{Gillett1973, Whittet1983, Gibb2004, Oberg2011}. \ce{H2O}, \ce{CO2}, and CO are the most abundant molecular ices in protostars, with lower abundances of \ce{CH3OH}, \ce{NH3}, \ce{CH4}, and OCN$^{-}$ \citep{Gibb2004, Oberg2011, Boogert2015}. Processing of these simple species by high-energy particles, radiation, and thermal heating is expected to lead to the build-up of larger, complex molecules.  Previous generations of IR telescopes, including the Infrared Space Observatory (ISO) \citep{Schutte1999, Gibb2004} and Spitzer \citep{Boogert2008, Oberg2011}, hinted at the presence of complex organic molecules in the ice phase during these early evolutionary stages. The high sensitivity and spectral resolution of the James Webb Space Telescope (JWST) has recently enabled deeper observations of ice absorption features in protostars than previously possible, confirming the presence of a diverse and chemically complex ice inventory \citep{Yang2022,McClure2023, Rocha2024, Brunken2024, Chen2024, Slavicinska2024, Brunken2025, Gross2025, Slavicinska2025, Rayalacheruvu2025, Rocha2025, EvD2025}.  Quantifying the abundances of simple and complex ice species in protostars is key to understanding the solid-phase volatile reservoir available at the onset of planet formation.

At near- to mid-infrared wavelengths, a protostar's spectrum may be influenced by (scattered) emission from the central protostar, thermal emission from dust grains, and absorption features from solids including ices and silicates.  Careful spectral analysis is required for the detection and quantification of ice absorption features. Common practice is to first produce a model of the continuum flux ($F^{cont}_\lambda$), by fitting a polynomial to ice-free wavelength regions \citep{Boogert2008}.  Then, the observed spectrum can be converted from flux units ($F^{source}_{\lambda}$) to optical depth ($\tau_\lambda$) using the equation: 

\begin{equation}
\tau_\lambda = -\mathrm{ln}(\frac{F^{source}_{\lambda}}{F^{cont}_{\lambda}})
\label{odequation}
\end{equation}

This assumes that the continuum model accurately represents the background flux which is attenuated by the absorbing medium in the foreground. Next, the optical depth due to silicate absorption must be accounted for in order to isolate the contributions of ice features. Silicate features can be removed by either assuming a common silicate profile like Galactic Center Source 3 (GCS 3) \citep{Kemper2004}, or by computing silicate opacities by assuming a grain size distribution, carbon fraction, and silicate type \citep{Rocha2024}. After subtracting the silicate profile, the optical depth spectrum that remains is assumed to be entirely due to ice, and is then typically fitted with a linear combination of reference laboratory ice data. Column densities of molecules can then be calculated from the integrated optical depth of each absorption band.

Although this approach is computationally efficient, it introduces several major uncertainties in the derived ice column densities.  Previous analysis has shown that the choice of the continuum model used to create an optical depth spectrum can strongly impact the optical depth of the ice absorption bands, in some cases dominating the uncertainty in the calculated column density \citep{Keane2001, Rocha2025, Gross2025}.  Subtraction of the silicate features also involves numerous assumptions about the grain properties, and neglects the mutual influence of silicate and icy material on the overall solid optical properties. Additional radiative transfer effects can also alter the observed absorption features; for instance, grain growth can induce scattering wings, which are distortions in the absorption band profile on the red (long-wavelength) or blue (short-wavelength) side.  In addition, mixing of light from multiple regions may be important if the IR source is spatially extended \citep{Pontoppidan2005, Sturm2023, Dartois2024}. 

These issues can be circumvented by self-consistently modeling the dust continuum, silicate absorption, and ice absorption features using radiative transfer models.  Indeed, there are numerous studies investigating these effects for mature protoplanetary disks \citep{Pontoppidan2005,Arabhavi2022,Dartois2022,Sturm2023,Bergner2024}.  Prior radiative transfer modeling of protostellar envelopes has included the presence of ice absorption features \citep[e.g.][]{Osorio2003,Crapsi2008,Furlan2008,Poteet2011}, but these works largely focused on using the models to constrain the envelope physical structure and silicate properties.  We are therefore still lacking a clear understanding of how the radiative transfer through icy protostellar envelopes impacts the ice spectral features observed at IR wavelengths.  This is a major limitation for our ability to both (i) accurately quantify the ice abundances in protostars, and (ii) interpret what material is actually being traced by the observed absorption features.

In this paper, we showcase a new radiative transfer framework designed to model ice absorption spectra of embedded protostars.  We apply this framework to develop a tailored model of the protostar IRAS 15398-3359, making use of JWST observations taken as part of the CORINOS program, (``COMs ORigin Investigated by the Next-generation Observatory in Space'', ID 2151, P.I. Y.-L. Yang).  Section \ref{sec:obs} presents the observations and Section \ref{sec:model} introduces the modeling framework.  In Section \ref{sec:results} we present our best-fitting model and describe the ice and grain properties inferred.  In Section \ref{sec:discussion} we discuss the implications for determining the optical depths of ice absorption bands and for interpreting what regions of the envelope are traced by the observed ice absorption features.

\section{Target Source and Observational Data}
\label{sec:obs}

IRAS 15398-3359 (hereafter IRAS 15398) is a very low-mass, Class 0 protostar located in the Lupus I molecular cloud at a distance of 154.9 $\pm$ 3.4 pc \citep{Heyer1989, Okoda2018, Galli2020}. A compact, embedded protostellar disk around 4 au in size has been associated with IRAS 15398 by \cite{Thieme2023}, however the disk's properties remain poorly constrained. By fitting the \ce{C^18O} emission from 0.5'' ALMA observations, \cite{Yen2017} estimated a centrifugal radius of 20 $^{+50}_{-20}$ au. \cite{Okoda2018} inferred a rotationally supported disk with a centrifugal barrier of 40 au and centrifugal radius of 80 au using high-resolution (0.2'') ALMA observations. Adding to the differences in proposed disk radii, Keplerian motion has not been clearly established, leaving the nature of the central object uncertain.

The chemistry of IRAS 15398 has been extensively studied.  Millimeter observations have revealed few gas-phase emission lines from interstellar complex organic molecules (iCOMs), though \ce{CH3OH} and \ce{CH3OCHO} were detected recently, indicative of a possible hot corino \citep{Okoda2023}. The source does, however, have a rich carbon-chain chemistry that may be driven by a recent accretion burst \citep{Sakai2009, Jorgensen2013, Okoda2018, Okoda2023}. This is supported by the ring-like morphology of \ce{H^13CO+},  explained by the destruction of \ce{H^13CO+} by water vapor liberated from grains \citep{Jorgensen2013}. Mid-infrared observations of IRAS 15398 using Spitzer identified the ice species \ce{H2O}, \ce{CO2}, \ce{CH3OH}, \ce{CH4}, and \ce{NH3} \citep{Boogert2008, Pontoppidan2008, Oberg2008, Bottinelli2010}.

Recent observations of IRAS 15398 were conducted using MIRI-MRS on-board JWST as part of the CORINOS program, leading to the detection of common ice species (\ce{H2O}, \ce{CO2}, CO, \ce{CH3OH}, \ce{CH4}, \ce{NH3}), absorption features attributed to complex organic molecules, and gas-phase emission tracing disk and outflow structures \citep{Yang2022, Salyk2024, Okoda2025}.  In this work we make use of the CORINOS MIRI spectra, along with NIRSpec-IFU and NIRSpec-PRISM data from programs 1854 (PI: M.~McClure) and 6161 (PI: K.~Slavicinska), respectively.  1D spectra were extracted from each data cube with a $1''$ diameter aperture centered on the source. We also compare our model to the photometric measurements of IRAS 15398 listed in Table \ref{tab:photometry}.

\section{Modeling}
\label{sec:model}

\begin{figure*}

\centering
    \includegraphics[width=0.8\paperwidth]{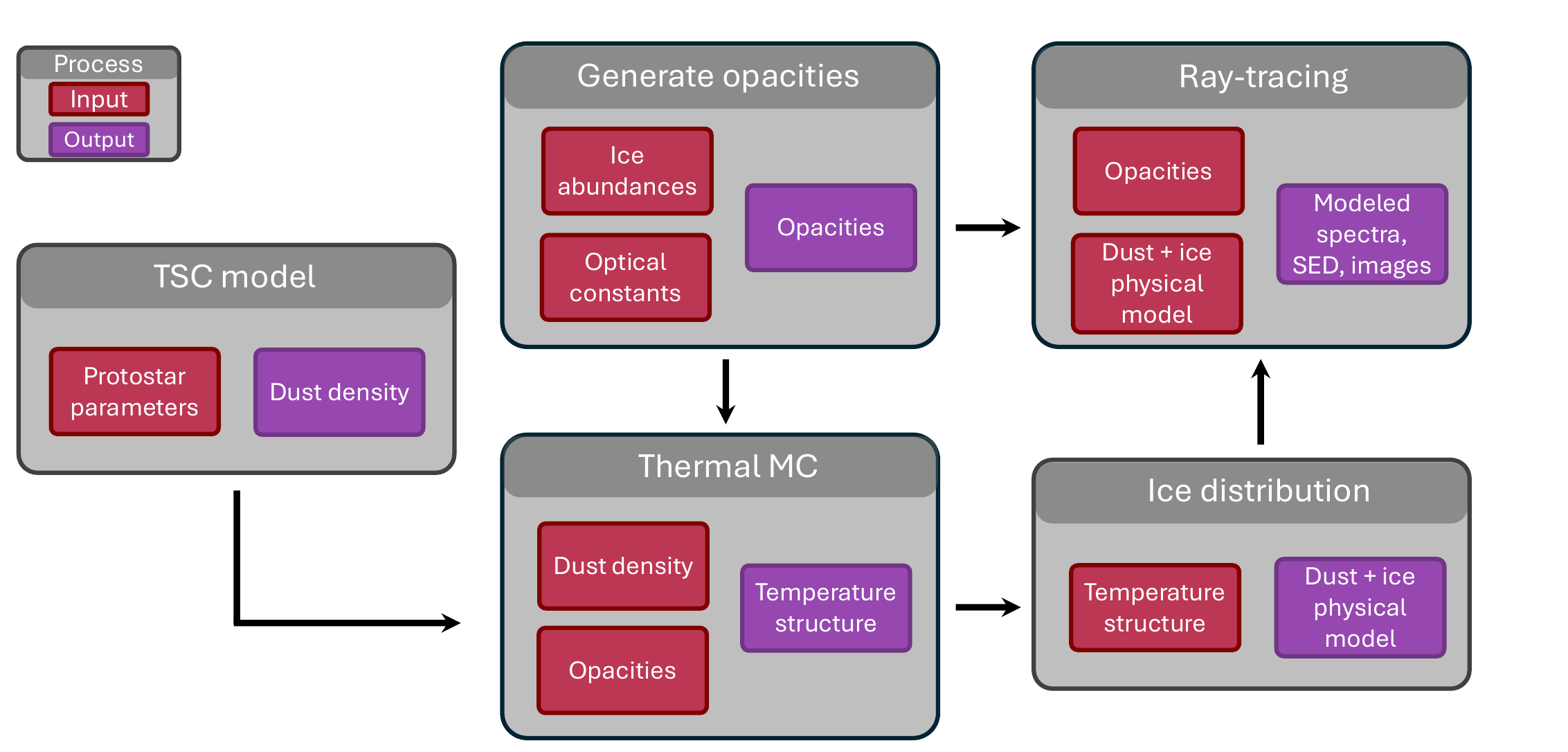}
    
    \caption{Schematic of the modeling framework, with model inputs (red), model outputs (purple), and processes (gray).}
    \label{fig:flow}

\end{figure*}

A schematic of the modeling framework is shown in Figure \ref{fig:flow}.  We begin with a physical model of the protostar and surrounding envelope based on the model described in \citet{Terebey1984} and \citet{Yang2017,Yang2020}.  The thermal Monte Carlo routine in \texttt{RADMC-3D} \citep{Dullemond2012} is used to calculate the dust temperature structure, which in turn sets the freeze-out zones of different ice species.  The dust opacities within each freeze-out zone are updated to account for the presence of ices, and then synthetic images are ray-traced at the wavelengths of interest.  All dust opacity calculations are carried out using the \texttt{jody} code \citep{Pontoppidan2024}, which itself makes use of \texttt{OpTool} \citep{Dominik2021}.  Additional detail on each step in the model is provided below.

\subsection{Physical Model}

\begin{figure*}
    \centering
    \includegraphics[width=0.85\paperwidth]{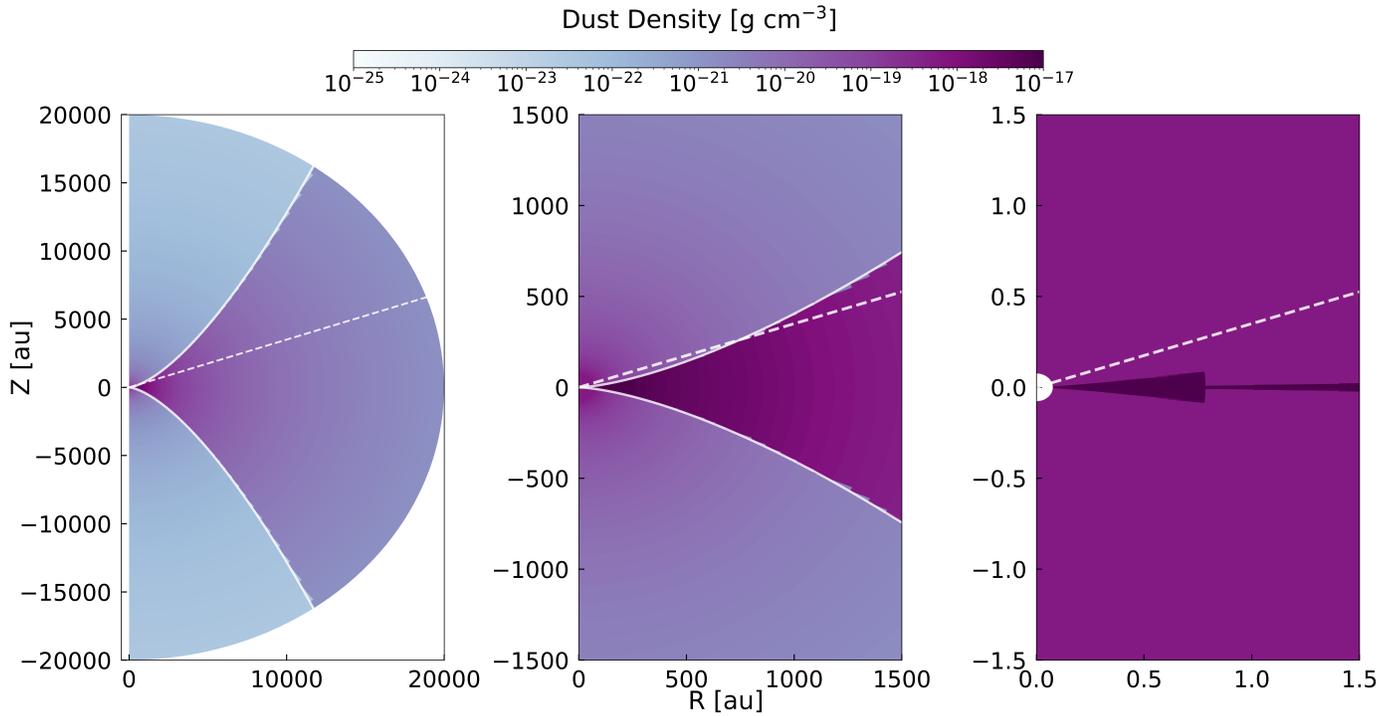}
    \caption{Dust density distribution in our preferred model of IRAS 15398. The white dashed line visualizes the pencil-beam line of sight based on the inclination of IRAS 15398 (71$^{\circ}$), and the white contour displays the cavity-envelope boundary in the model.}  From left to right, the panels are increasingly zoomed in around the center of the protostar. 
    \label{fig:2ddustdensity}
\end{figure*}

The protostellar envelope density structure is calculated with a slowly rotating collapse model, hereafter referred to as the TSC model \citep{Terebey1984}, which is also described in detail in \cite{Yang2017}. The TSC model calculates the envelope density structure based on the age, effective sound speed (c$_{s}$), and the initial rotation speed ($\Omega_o$). The two-dimensional ($r, \theta$) axisymmetric dust density distribution is given by Equation 4 in \cite{Yang2017}, from \cite{Ulrich1976}.  The outflow cavity was parameterized by an opening angle, $\theta_{cavity}$; a cavity power-law index, $\alpha_{cavity}$; a cavity radius, R$_{cavity}$; and a cavity density, $\rho_{cavity}$. The outer radius of the envelope was adopted as 20,000 au, sufficiently large that the results are insensitive to the inclusion of further radii.

A flared accretion disk was also added to the model based on \cite{Shakura1973} and \cite{Robitaille2006}, described by a mass ($M_{disk}$), scale height at 100 au (h$_{100}$), and flaring index ($\beta$), which describes the scale height as a function of radius $\varpi$ in cylindrical coordinates: $h \propto \varpi^{\beta}$. Details on the disk prescription can be found in \cite{Yang2017}. The disk inner radius corresponds to the calculated dust sublimation temperature (assuming a sublimation temperature of 1600 K), and the disk outer radius was set to the centrifugal radius calculated from the TSC model.

To illustrate the envelope model, Figure \ref{fig:2ddustdensity} shows the 2D dust density of the IRAS 15398 envelope for our preferred model.  The outflow cavities are apparent as low-density cones projecting vertically away from the center.  Note that the disk is apparent on the right panel of Figure \ref{fig:2ddustdensity} with a radius of 0.78 au.

The protostar itself is assumed to be a blackbody characterized by an effective temperature $T_{star}$ and radius $R_{star}$, which together set the simulated bolometric luminosity.  The envelope temperature structure is calculated using the thermal Monte Carlo calculation in RADMC-3D, based on the stellar parameters, envelope density structure, and dust opacities for ice-free grains. Viscous heating is not included in the temperature calculation.  Dust grain opacities for the thermal Monte Carlo calculations were calculated as described in Section \ref{sec:opacities}, but including no ice mantle component. $1.5\times10^{7}$ photons were used for this calculation. The calculated temperature structure is shown in Figure \ref{fig:2ddusttemp} in Appendix \ref{appendixdusttemp}. 

\subsection{Ice model}
We focus on modeling H$_2$O, CO$_2$, and CO ices since these species are sufficiently abundant that their presence on grains can influence the overall shape of the silicate continuum (e.g.,~Figure \ref{fig:iceeffect}).  Their inclusion is therefore necessary to self-consistently model the protostellar physical structure and dust and ice properties.  In order to keep the model fitting tractable, we chose not to include additional ice species which are likely present at lower abundances, like CH$_3$OH and NH$_3$. Indeed, a major goal of this study is to assess how more realistic radiative transfer influences the calculation of ice-phase column densities (Section \ref{sec:discussion}).

In the model, a given molecule is assumed to freeze onto grains at locations within the protostellar envelope where the dust temperature is less than its freeze-out temperature.  The freeze-out temperatures are calculated according to \citet{Hollenbach2009}, which equates the steady-state flux of molecules adsorbing and desorbing from grains:
\begin{equation}
    T_f = E_b\Bigg{[}57 + \mathrm{ln}\Big{(}\frac{1 \;\mathrm{cm^{-3}}}{n_{gas}}\Big{)} \Big{(}\frac{10^4 \;\mathrm{cm\;s}^{-1}}{v_i}\Big{)}\Bigg{]}^{-1},
\end{equation}
where $E_b$ is the binding energy, $n_{gas}$ is the gas density, determined here assuming a local gas-to-dust ratio of 100, and $v_i$ is the thermal speed for a given molecule, equal to $\sqrt{k_BT/m_i}$, where $k_B$ is the Boltzmann constant, $T$ is the local dust temperature in a given grid cell, and $m_i$ is the mass of the molecule.  The binding energies used for calculating freeze-out zones are listed with other molecular constants in Table \ref{constants}.  

\begin{figure}
    \centering
    \includegraphics[width=0.98\columnwidth]{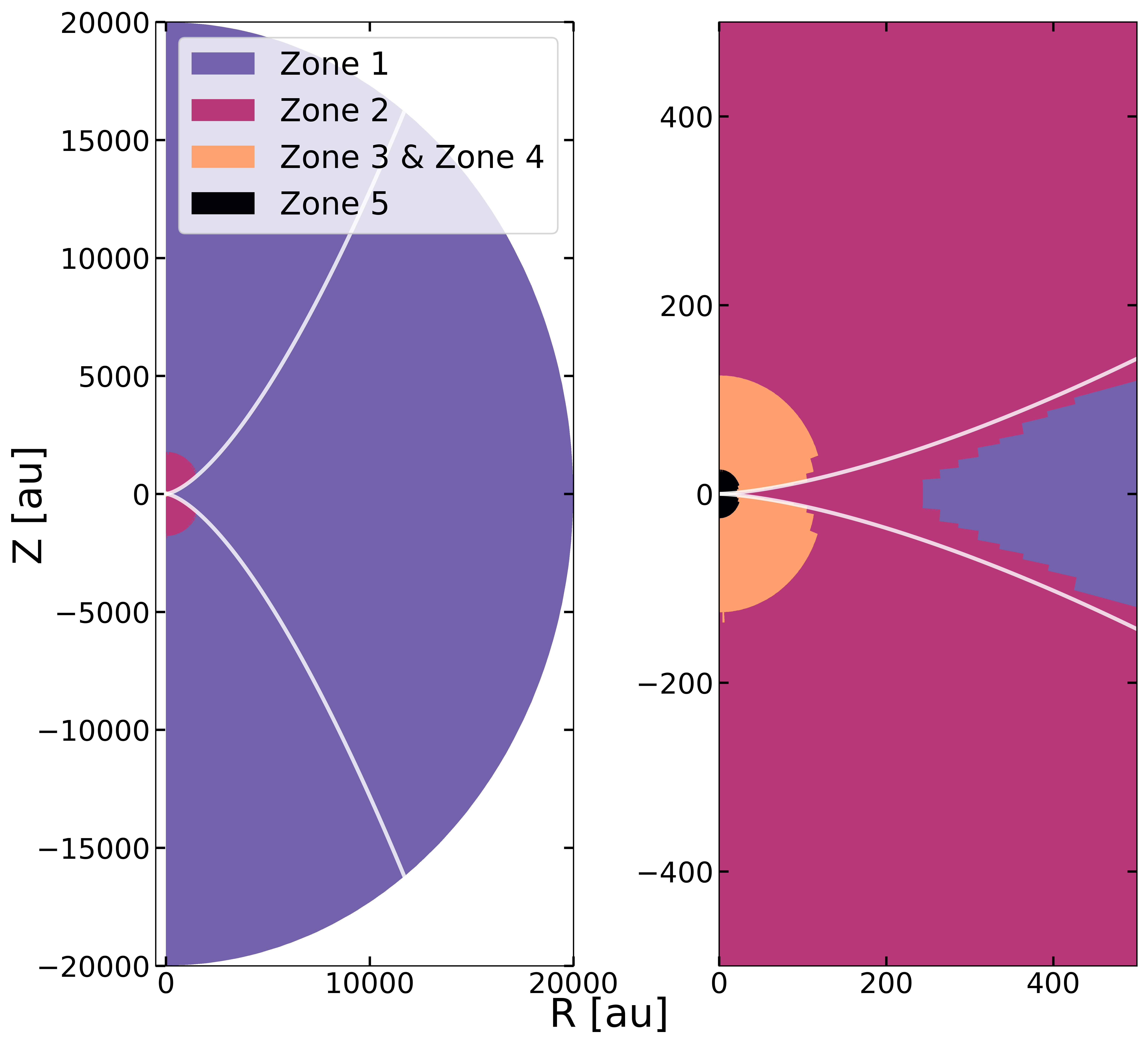}
    \caption{Envelope zones for our preferred model of IRAS 15398, determined from the freeze-out boundaries of H$_2$O, CO$_2$, and CO. The right panel shows a zoom-in of the inner 500 au. Zones are defined in Table \ref{tab:dust_zones}. Zones 3 \& 4 (defined by the amorphous + crystalline water freeze-out boundaries) were combined for clarity. The white contour displays the cavity-envelope boundary in the model.} 
    \label{fig:dustzones}
\end{figure}

Based on the freeze-out boundaries of each molecule in our model, the envelope is then divided into different zones.  Dust opacities are calculated for each zone according to what ices are present.  Figure \ref{fig:dustzones} illustrates the different envelope zones for our preferred physical model of IRAS 15398.  Table \ref{tab:dust_zones} lists the solid species contained within each zone, and the labeling convention used hereafter.  Zone 1 is furthest from the protostar, in the coldest regions where all ice species are frozen out, while Zone 5 is closest to the protostar in the warmest regions with no ice.  Note that amorphous and crystalline water are included as separate ice species in the model, but are combined for clarity in Figure \ref{fig:dustzones} since there is only a small difference in their spatial distribution.

\begin{table}[h!]
\centering
\caption{Envelope zones \label{tab:dust_zones}}
\begin{tabular}{c c c} 
\hline\hline
 Solid species & Label  \\ [0.5ex] 
 \hline
 Dust, \ce{H2O}a, \ce{H2O}c, \ce{CO2}, CO & Zone 1 \\ 
 Dust, \ce{H2O}a, \ce{H2O}c, \ce{CO2}     & Zone 2 \\
 Dust, \ce{H2O}a, \ce{H2O}c               & Zone 3 \\
 Dust, \ce{H2O}c                          & Zone 4 \\
 Dust                                     & Zone 5 \\
 \hline
\end{tabular}
\begin{tablenotes}
   \item Note: H$_2$Oa refers to amorphous water, and H$_2$Oc to crystalline water.  CO$_2$ and CO are assumed to be fully amorphous.
\end{tablenotes}

\end{table}

\subsection{Opacities}
\label{sec:opacities}
\begin{figure*}[h]
    \centering
    \includegraphics[width=0.85\paperwidth]{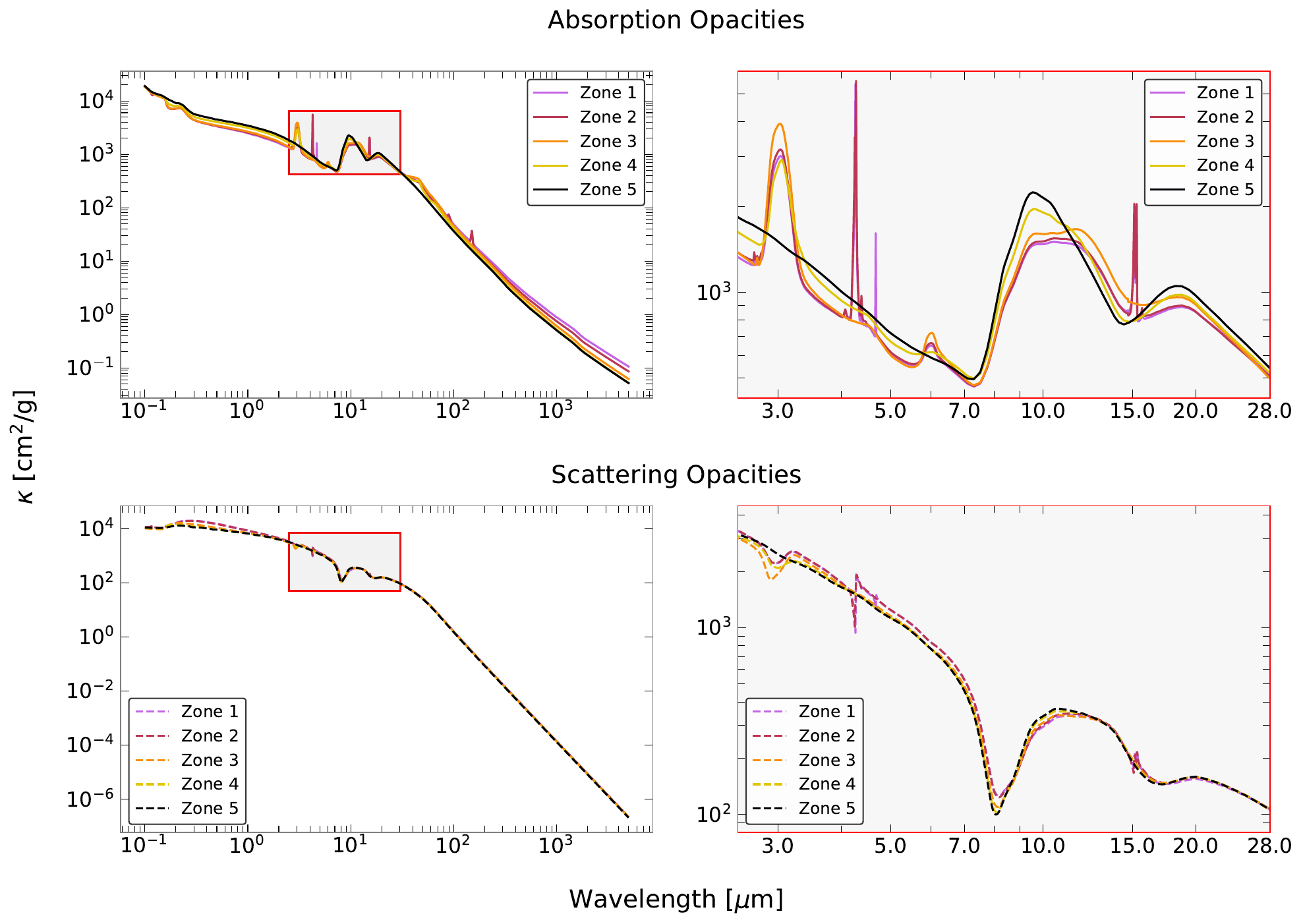}
    \caption{Absorption (top row) and scattering (bottom row) opacities for each envelope zone in our preferred model of IRAS 15398.  Zone 1 contains all ice species, and Zone 5 contains only refractory dust.  The right columns show a zoom-in of the 2.5 -- 28 $\mu$m range, outlined with red boxes in the left column.} 
    \label{fig:opacities}
\end{figure*}

Dust opacities are calculated using the \texttt{jody} code described in \cite{Pontoppidan2024}, which uses the OpTool framework to calculate absorption and scattering cross-sections \citep{Dominik2021}. \texttt{jody} assumes two distinct populations of the grain `cores', one refractory carbon and one silicate.  Each population is characterized by its own grain size distribution adapted from \cite{Weingartner2001}, parameterized by a minimum and maximum size, $a_{min}$ and $a_{max}$, and power-law index $\alpha$.  $c_{frac}$ denotes the mass fraction of carbon grains relative to the total mass of carbon + silicate grains.  The silicate and carbon grain opacities were calculated using the `astrosilicate' (\ce{MgFe(SiO)4}) and amorphous carbon species within OpTool, based on the optical constants from \cite{Draine2003} and \cite{Zubko1996}. The opacity calculation used the Distribution of Hollow Spheres method, which is suitable for irregular and low-porosity aggregates and has been shown to reproduce the scattering, absorption, and emission from cosmic dust grains \citep{Min2005}.

Absorption and scattering opacities were calculated for each envelope zone to account for the presence of different ices.  Ices are included as `mantle' components coating the refractory cores, which had a maximum thickness of 0.14 $\mu$m.  The amorphous water ice abundance is parameterized relative to hydrogen, assuming a gas-to-dust ratio of 100, while other ice species are parameterized as fractional abundances with respect to amorphous water.  The ice mantles are assumed to have a constant thickness across all grains with sizes ($a$) $>$ 0.03 $\mu$m within a given population; grains with $a$ $<$ 0.03 $\mu$m are treated as ice-free due to transient heating \citep{Pontoppidan2024}. The ice mantle opacities were calculated using optical constants from \cite{Hudgins1993} (\ce{H2O}a), \cite{Warren2008} (\ce{H2O}c), \cite{Warren1986} (\ce{CO2}), and \cite{Palumbo2006} (CO).  The opacities for our preferred model of IRAS 15398 are shown in Figure \ref{fig:opacities}.

\subsection{Comparison to observations}
\label{subsec:fitting}
Based on the model dust distribution and dust opacities, synthetic image cubes were ray-traced in \texttt{RADMC-3D} using $1.5\times10^{7}$ photons per wavelength and the full treatment of anisotropic scattering with polarization.  The wavelength sampling for the synthetic cubes was chosen by hand to ensure that all continuum and ice features are spectrally resolved.  Images were convolved with a 2D gaussian determined by the diffraction limit, $\sim$1.22$\lambda$/D, at each wavelength.  1D spectra were extracted with a 1'' diameter aperture centered on the source, analogous to how the observed spectra were extracted.  We also produced a synthetic spectral energy distribution by ray-tracing at the wavelengths of observed photometry and extracting fluxes using the appropriate aperture sizes (Table \ref{tab:photometry}).

Given the high computational expense of the ray-tracing and the large number of parameters in the  model (16 physical parameters + 6 grain parameters + 4 ice parameters), a statistical fitting of our model to the observations is not tractable.  The model was instead tuned by hand to reproduce the observed JWST near- and mid-IR spectra along with the spectral energy distribution.  The wavelength ranges from $\sim$3.5--4$\mu$m and 5.5--10$\mu$m contain absorption features from species not included in our  model (e.g.~hydrates, methanol, NH$_3$, larger organics) and we do not attempt to match these regions of the spectrum, instead focusing on the continuum and absorption features from H$_2$O, CO$_2$, and CO.

The source age was first constrained by fitting the sub-millimeter radial profile, as described in Appendix \ref{appendixage}.  Next, the age was fixed and initial guesses for the protostellar physical structure were informed by modeling the SED alone.  After this, several physical parameters were fixed for the fine-tuning stage: effective sound speed, initial angular speed, dust-to-gas ratio.  Additionally, we fixed the disk parameters $\beta$ and $h_{100}$ as they did not noticeably influence the resulting spectra.  This reduced the number of free parameters for the physical structure to 9.  A detailed assessment of parameter degeneracies and uncertainties in this model is forthcoming in Kreft et~al. (in prep).

\section{Results}
\label{sec:results}

\subsection{Comparison of preferred IRAS 15398 model to observations}
\label{subsec:model_bestfit}

\begin{deluxetable}{lll}
	\tablecaption{IRAS 15398 model parameters \label{tab:params}}
	\tablecolumns{3} 
	\tablehead{
        \colhead{Parameter}        & 
        \colhead{Description} & 
        \colhead{Value}      }
\startdata
\multicolumn{3}{c}{\textit{Envelope \& Outflow Parameters}} \\
T$_{star}$ & Stellar temperature & 4000 K \\
R$_{star}$ & Stellar radius & 5.25 R$_{\odot}$ \\
R$_{disk}$ & Disk radius & 0.78 au \\
M$_{disk}$ & Disk mass & $1\times10^{-4} M_{\odot}$ \\
$\theta_{inclination}$ & Viewing inclination & 71$^{\circ}$ \\
$\theta_{cavity}$ & Cavity opening angle & 44$^{\circ}$ \\
$\rho_{cavity}$ & Cavity density & 8.15 $\times 10^{-20}$ g cm$^{-3}$ \\
R$_{cavity}$ & Cavity radius  & 70 au \\
$\alpha_{cavity}$ & Cavity power law index & 1.5 \\
t & Age &  $1\times10^{4}$ yr \\
c$_s$ & Effective sound speed & $\bm{4\times10^{4}}$ cm s$^{-1}$ \\
$\Omega_{0}$ & Initial angular speed & $\bm{4\times10^{-13}}$ s$^{-1}$ \\
$\beta$ & Disk flaring index & $\bm{1.3}$ \\
h$_{100}$ & Scale height at 100 au & $\bm{8}$ au \\
$\epsilon$ & Gas-to-dust ratio & $\bm{100}$ \\
r$_0$ & Outer envelope radius & $\bm{20,000}$ au \\
\hline
\multicolumn{3}{c}{\textit{Grain Parameters}} \\
a$_{min}$ & Minimum grain size & 5$\times10^{-4}$ $\mu$m \\
a$_{max,si}$ & Silicate maximum size & 6.20 $\mu$m \\
a$_{max,c}$ & Carbon maxiumum size & 0.40 $\mu$m \\
$\alpha_{si}$ & Silicate power law & -2.00 \\
$\alpha_{c}$ & Carbon power law & -1.75 \\
c$_{frac}$ & Carbon fraction & 0.115 \\
\hline
\multicolumn{3}{c}{\textit{Ice Abundances}} \\
n\ce{H2Oa} & w.r.t. nH & 2.90$\times10^{-4}$ \\
n\ce{H2Oc} & w.r.t. n\ce{H2O}a & 0.25 \\
n\ce{CO2} & w.r.t. n\ce{H2O}a & 0.95 \\
nCO & w.r.t n\ce{H2Oa} & 0.25 \\
\enddata
\tablenotetext{}{Parameters in bold were held constant during modeling.}
\end{deluxetable}

\begin{figure*}[ht!]
    \centering
    \includegraphics[width=0.85\paperwidth]{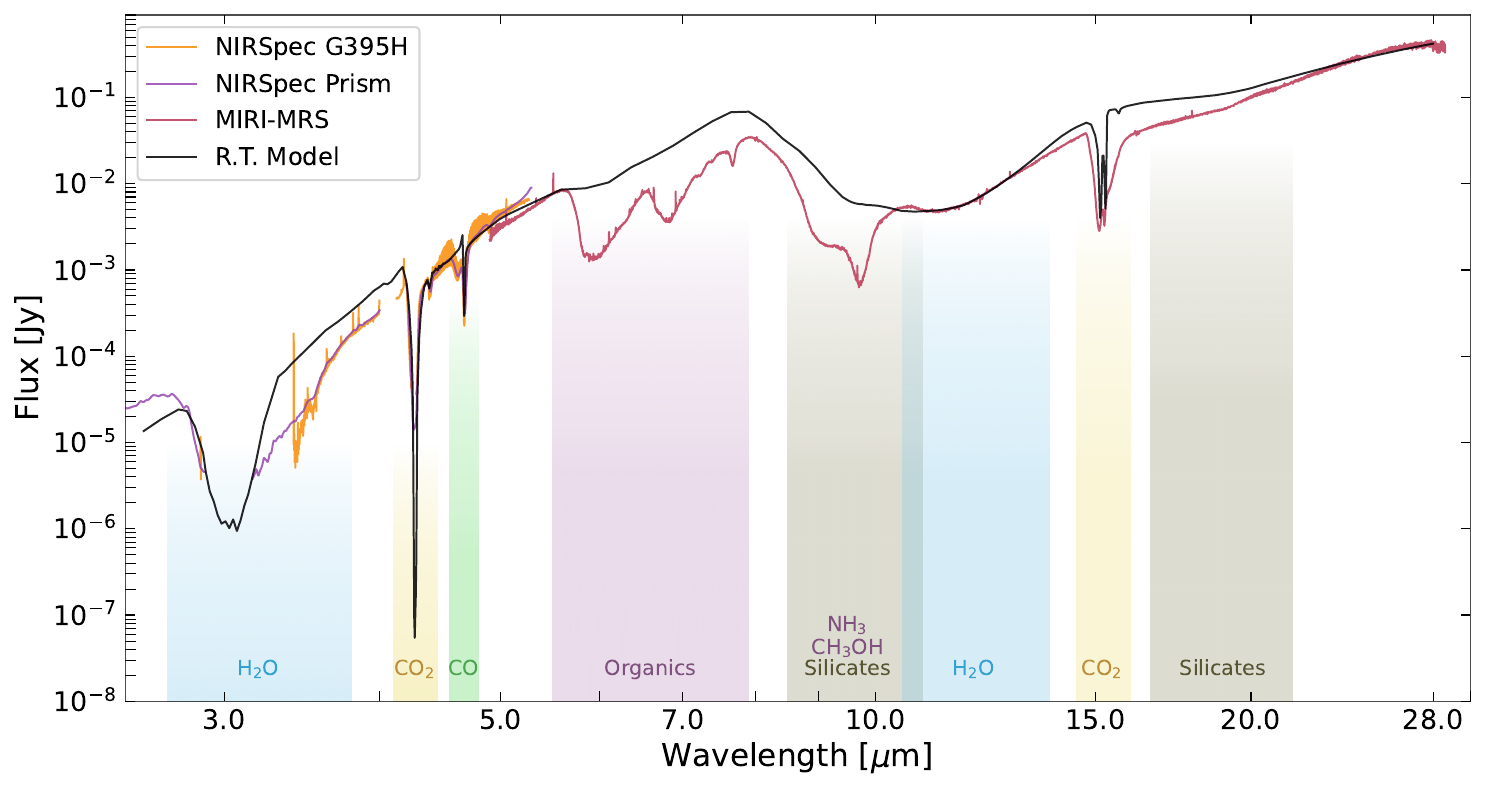}
    \caption{IRAS 15398 JWST NIRSpec spectrum (Program ID 1854, PI: M. McClure), NIRSpec Prism spectrum (Program ID 6161, PI: K. Slavicinska), and MIRI-MRS spectrum (Program ID 2151, PI: Y.-L. Yang) with the radiative transfer model spectrum extracted in a 1'' diameter overlaid in black. Data points which hit the noise floor from the G395H and Prism data were removed. A spectrum zoomed in to the NIR \ce{CO2} and CO absorption is shown in Appendix \ref{appendixRTzoom}. } 
    \label{fig:spectrum}
\end{figure*}

Figure \ref{fig:spectrum} shows the near- and mid-IR spectrum for our preferred model of IRAS 15398, compared to the JWST observations.  Table \ref{tab:params} reports the corresponding model parameters.  This model provides an excellent match to the overall IR continuum as well as the absorption features from silicates, H$_2$O, CO$_2$, and CO ice.  Note that the 3~$\mu$m water ice feature hits the noise floor of the data, making it difficult to constrain the water abundance based on this band alone, and underscoring the importance of simultaneous fitting of the 12~$\mu$m libration mode.  Reproducing the shape of the water libration band required the inclusion of both amorphous and crystalline water in the model.  The 4.5~$\mu$m CO$_2$ stretch is also saturated, and the 15.5~$\mu$m bending mode served as our primary anchor of the CO$_2$ ice abundance.  The $^{13}$CO$_2$ stretch provided a secondary check of the adopted CO$_2$ abundance.

The model and observations deviate in the $\sim$3.5--4~$\mu$m and 5.5--10~$\mu$m ranges, as expected due to the absorptions from ice species that were not included in our model (Section \ref{subsec:fitting}).  In Section \ref{sec:tau_compare} we use our modeled continuum to determine how the optical depths of these features compare to a continuum model using the polynomial method.

\begin{figure}
    \centering
    \includegraphics[width=0.90\columnwidth]{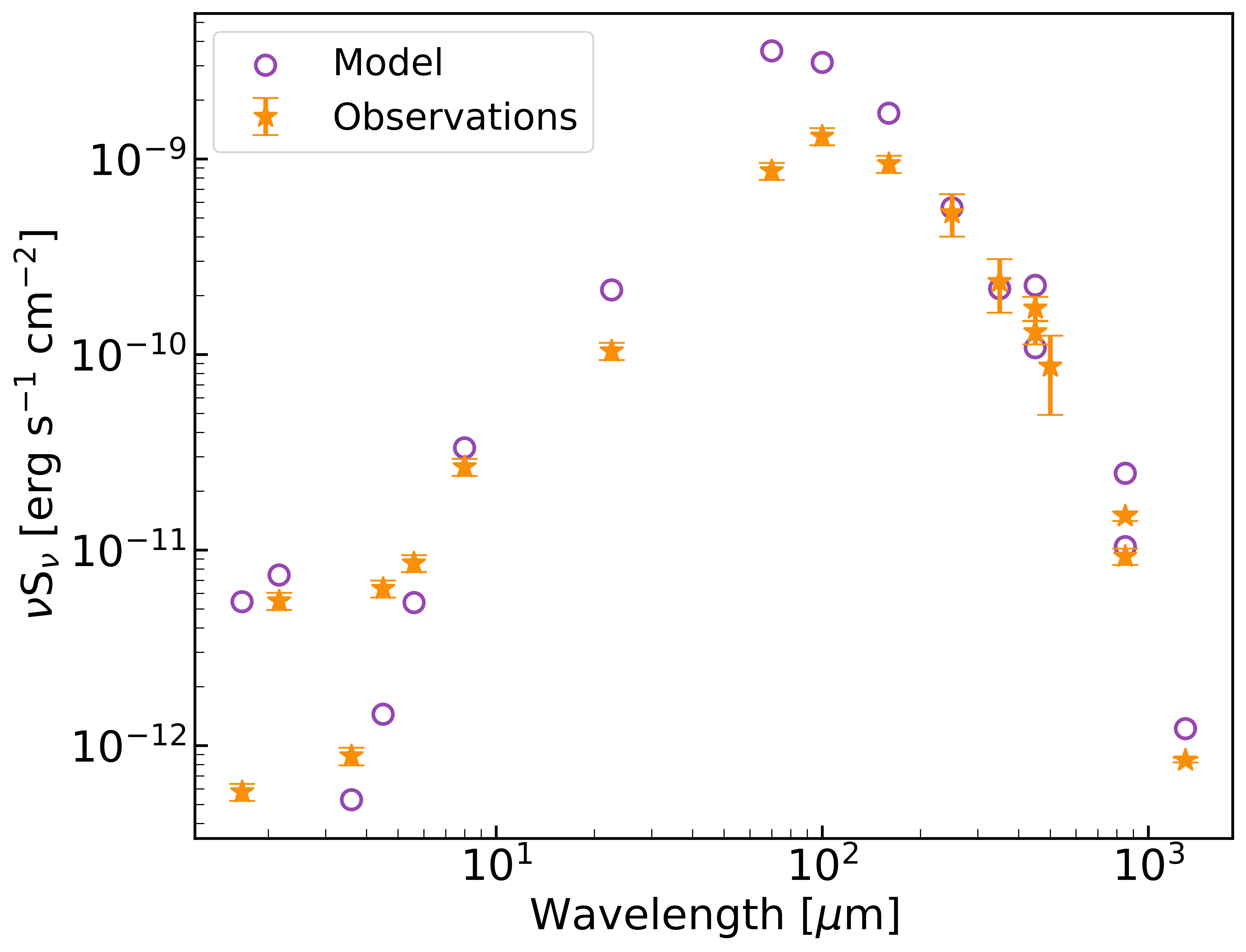}
    \caption{Comparison of the observed spectral energy distribution (orange stars) with our preferred IRAS 15398 model (purple circles).} 
    \label{fig:sed}
\end{figure}

The modeled vs.~observed spectral energy distribution is shown in Figure \ref{fig:sed}.  The synthetic SED is overall in good agreement with the observed photometry, especially from 250 -- 1300 $\mu$m.  The largest deviations between the modeled and observed fluxes occur around 70--160$\mu$m, possibly due to flux variability in the source. 
The model also deviates somewhat from the observed fluxes in the near- and mid-IR range.  However, given the excellent agreement with the JWST spectrum, this discrepancy is likely due to the larger apertures used to extract SED fluxes.  Our model therefore describes the inner protostellar envelope well, but may not fully capture the larger-scale structure.

\subsection{Inferred protostar physical properties}
The bolometric luminosity calculated from integrating the simulated SED, given a stellar effective temperature of 4000 K and stellar radius of 5.25 R$_{\odot}$, is 5.29 L$_{\odot}$. Previous observations with Herschel-SPIRE found an apparent bolometric luminosity of 1.49 L$_{\odot}$. 

The modeled inclination of 71$^{\circ}$ aligns with the value of 66 $\pm$ 14$^\circ$ found by \cite{Vazzano2021} and the value of 70$^\circ$ adopted by \cite{Okoda2018}.  Note that the lower inclination angle of 50.7$^\circ$ estimated by \cite{Thieme2023} is based on the disk whereas ours and that of \cite{Okoda2018} are constrained by the outflow, and it is possible that there is a misalignment in the system between the small and large scales.  The outflow-based inclination is more relevant for our analysis since this determines the extent to which the line of sight passes through the cavity (Section \ref{subsec:ice_origins}). The disk radius of 0.78 au is smaller than the $\sim$4 au dust disk found by \cite{Thieme2023}, who calculated the dust disk radius by taking the 2$\sigma$ radius of 2D Gaussian fits of the continuum. Note that in our model, the presence of a somewhat larger disk does not influence the synthetic spectrum and therefore is not ruled out.  Our modeled cavity opening angle of 44$^{\circ}$ is somewhat higher than the value of $\sim$30$^\circ$ found by \cite{Vazzano2021}, and was needed for our models to correctly reproduce the shape of the JWST MIRI-MRS spectrum.  We find an age of $1\times10^{4}$ yr which is somewhat younger than the value of $3\times10^{4}$ yr determined by fitting radial profiles of the sub-millimeter flux \citep{Shirley2000}. 

\subsection{Inferred ice and dust properties}
The ice abundances inferred from our preferred model are listed in Table \ref{tab:params}.  We find a total water abundance of 3.6$\times$10$^{-4}$ H$^{-1}$, of which 80\% is amorphous and 20\% is crystalline.  This corresponds to $\sim$70\% of the volatile + refractory cosmic oxygen abundance of $5\times10^{-4}$ from \cite{Meyer1998}, though note that there is degeneracy between the absolute ice abundances and the assumed gas-to-dust mass ratio.  This water abundance is considerably higher than that of \textit{KP5} from \cite{Pontoppidan2024}, which used a water abundance of 7.3$\times10^{-5}$ H$^{-1}$. CO$_2$ has a fractional abundance of 76\% with respect to the total (amorphous + crystalline) water abundance; this is higher than what is typically retrieved towards protostars.  CO has a fractional abundance of 20\% with respect to the total water abundance.

\begin{figure}
    \centering
    \includegraphics[width=0.90\columnwidth]{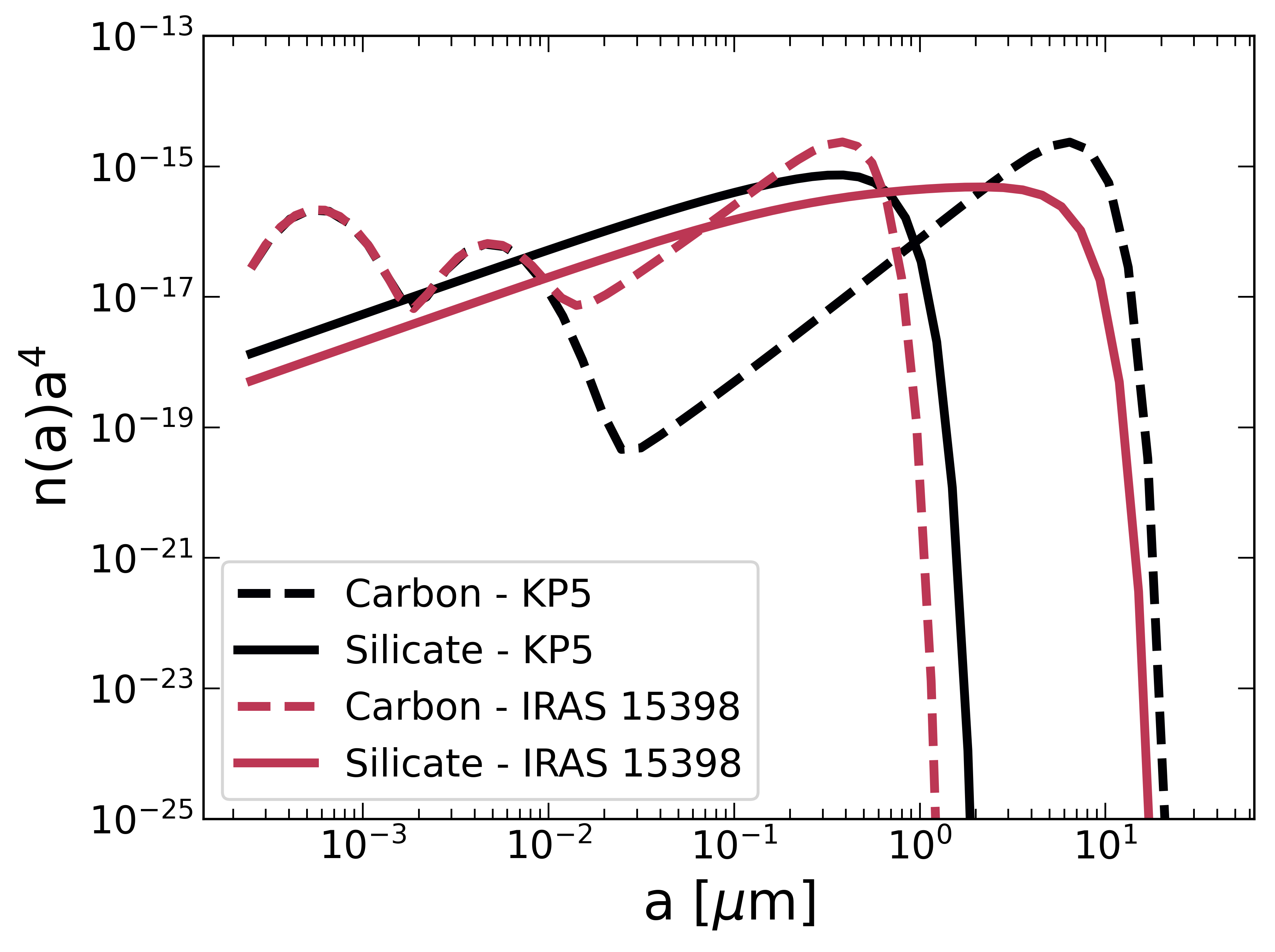}
    \caption{Size distribution of the separate carbon and silicate grain populations in the preferred model of IRAS 15398. The $KP5$ model for dense clouds presented in \citet{Pontoppidan2024} is shown for comparison.}
    \label{fig:graindist}
\end{figure}

Figure \ref{fig:graindist} shows the size distribution of carbon and silicate grain populations in our preferred model; for comparison we also show the \textit{KP5} model presented in \citet{Pontoppidan2024}, which was constrained by dense cloud observations.  Our preferred model uses a minimum grain size of $5\times10^{-4}$~$\mu$m for both the silicate and carbon grain populations, though the spectrum is not very sensitive to changes in this parameter below $\sim$10$^{-3}$~$\mu$m.  The carbon grains account for 11.5\% of the total refractory grain mass; this was constrained mainly by the slope of the modeled spectrum from 3 - 8 $\mu$m, as well as the depth of the 10 $\mu$m silicate absorption feature. Note that the carbon grain population includes very small size polycyclic aromatic hydrocarbons (PAHs) and larger graphitic grains, similar to the carbon grains modeled in \cite{Li2001}; this gives rise to the multiple distinct peaks in the grain size distribution. Unlike \textit{KP5}, our model prefers a larger maximum size for silicate grains (6.2~$\mu$m) compared to carbon grains (0.4~$\mu$m).  A larger carbon-grain size was found to flatten the overall spectrum, which did not fit the slope of the NIR and MIR observations. Neglecting the difference between the two different grain populations, our model has a comparable maximum grain size, $\sim$6--7~$\mu$m, as \textit{KP5}; we therefore do not see evidence for significant grain growth between the dense cloud and protostellar envelope stage. However, we primarily used the mid-IR wavelengths to constrain the grain size distribution, and with a uniform dust distribution across the entire envelope, our model may have limited ability to constrain the presence of larger grains in the denser inner envelope.

\subsection{Ice column densities}
\label{subsec:ice_columns}

\begin{table}[h!]
\centering
\footnotesize
\begin{threeparttable}
\caption{Calculated Column Densities ($\times10^{19}$ cm$^{-2}$)}
\begin{tabularx}{\columnwidth}{@{}
p{0.20\columnwidth}
p{0.05\columnwidth}
p{0.05\columnwidth}
p{0.05\columnwidth}
p{0.15\columnwidth}
p{0.15\columnwidth}
@{}}

\toprule
Source & \ce{H2O} & \ce{CO2} & CO & \ce{CO2}/\ce{H2O} & CO/\ce{H2O} \\
\midrule
IRAS 15398 (This work) & 1.95 & 1.41 & 0.36 & 0.72 & 0.18 \\
IRAS 15398$^{a}$ & 1.71 & 0.46 & -- & 0.27 & -- \\
IRAS 15398$^{b}$ & 1.48 & 0.52 & -- & 0.35 & -- \\
\hline
Typical low-mass protostar$^{c}$ & -- & -- & -- & 0.29 & 0.29 \\
\bottomrule
\end{tabularx}
\label{columndensitytable2}
\begin{tablenotes}
\item $a$: \cite{Kim2025}, $b$: 
\cite{Pontoppidan2008}, $c$: 
\cite{Oberg2011b}
\end{tablenotes}
\end{threeparttable}
\end{table}

From our preferred model structure, we calculated the  \ce{H2O}, \ce{CO2}, and CO ice column densities along the viewing line of sight to the central star, based on the dust mass and the ice-to-dust ratios within each envelope zone.  This approach provides the `true' column density in the model, which can be compared to the empirical column densities derived from observations, assuming that the observed column densities trace absorption against a point-source.  We confirmed that this is the case for the JWST observations of IRAS 15398, as the source size scales with wavelength as expected given the JWST PSF.  The modeled column density values are shown in Table \ref{columndensitytable2}, along with \textit{Spitzer} and JWST observations of IRAS 15398 and typical values measured in low-mass protostars.  Note that the \ce{H2O} column density in Table \ref{columndensitytable2} is the sum of the amorphous and crystalline components.    In addition to absolute column densities, we also show the CO$_2$/H$_2$O and CO/H$_2$O column density ratios.  Note that these are nearly the same as the inputted ice abundances (76\% CO$_2$/H$_2$O and 20\% CO/H$_2$O), discussed further in Section \ref{subsec:ice_origins}.

The modeled \ce{H2O} column densities are quite similar to those calculated empirically from the observed optical depths \citep{Pontoppidan2008, Kim2025}.  The relative CO column density is comparable to typical protostellar values of 20--40\% with respect to H$_2$O \citep{Oberg2011b}.  On the other hand, the CO$_2$ column density of 1.41 $\times10^{19}$ cm$^{-2}$ is 3$\times$ higher than what was previously found towards this source by \cite{Pontoppidan2008}.  The CO$_2$/H$_2$O column density ratio of 72\% is, similarly, higher than the range of $\sim$20--30\% typically determined for low-mass protostars via empirical optical depth analysis \citep{Oberg2011b}.

In tuning the model, we attempted several approaches to reduce the CO$_2$/H$_2$O abundance ratio (and in turn column density ratio) needed to reproduce the observations.  This includes testing different grain properties (like size distribution) and source properties (like inclination), using optical constants for CO$_2$:H$_2$O mixtures instead of pure CO$_2$, and adding a background radiation field.  In all cases, we still required a high CO$_2$/H$_2$O ratio close to unity.  

\begin{figure}
    \centering

    \includegraphics[width=0.90\columnwidth]{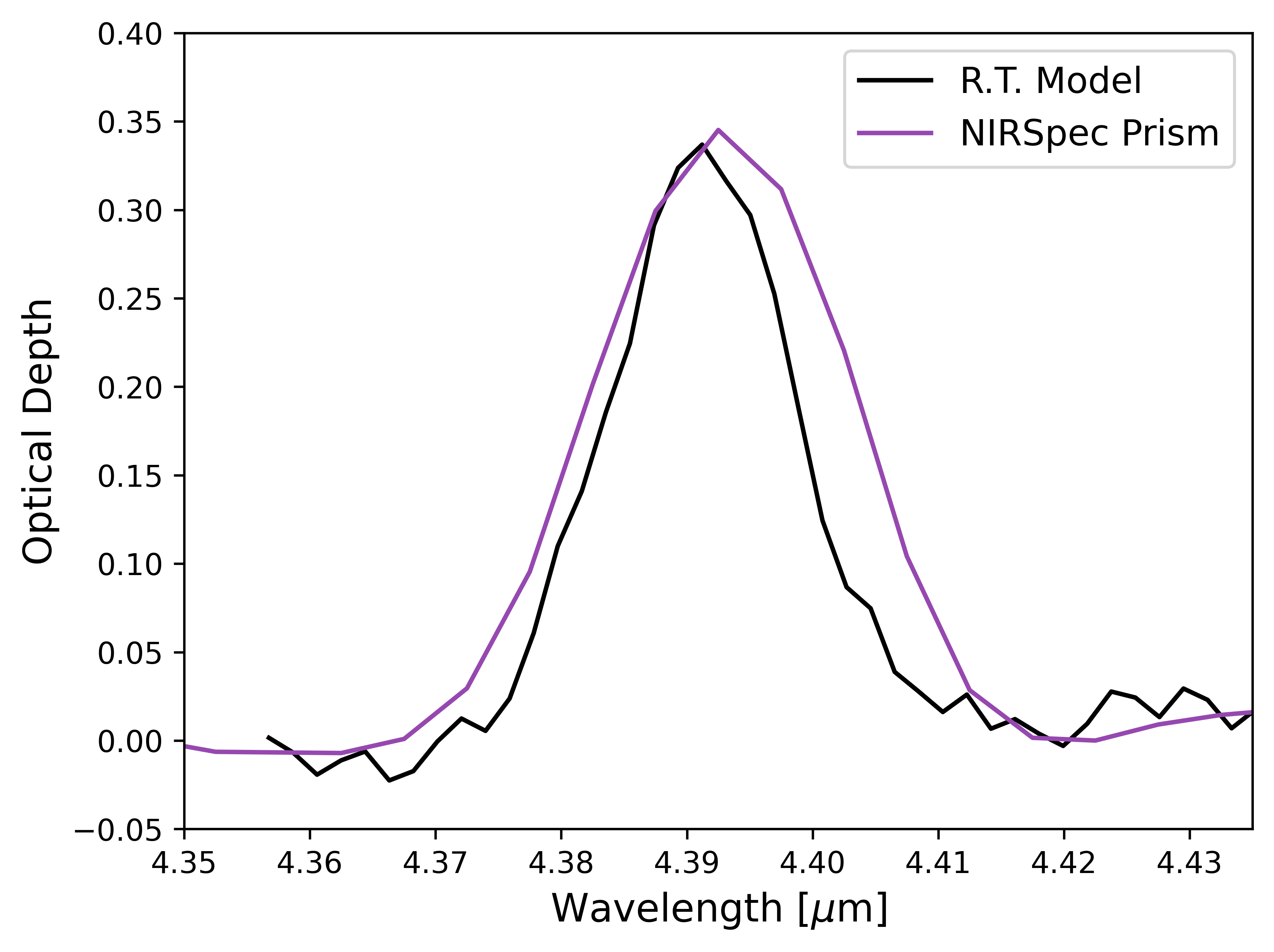}
    \caption{Optical depth comparison for modeled \ce{^13CO2} (black) and the NIRSpec Prism data (purple).  } 
    \label{fig:13co2}
\end{figure}

To cross-check the high inferred CO$_2$ ice abundance, we tested whether the optically thin $^{13}$CO$_2$ stretching band is reproduced for a reasonable $^{12}$C/$^{13}$C ratio.  Because there is some $^{13}$CO$_2$ absorption present in the optical constants of $^{12}$CO$_2$, to determine the $^{13}$CO$_2$ ice abundance we ran a separate model including only $^{13}$CO$_2$ and H$_2$O.  A $^{13}$CO$_2$ ice abundance of $8\times10^{-3}$ with respect to amorphous \ce{H2O} was found to reproduce the depth seen in the NIRSpec Prism observations (Figure \ref{fig:13co2}).  This corresponds to a \ce{^12CO2}/\ce{^13CO2} abundance ratio of 118. This is comparable to the reported mean \ce{^12CO2}/\ce{^13CO2} ratio of 97 $\pm$ 17 from \cite{Brunken2024, Brunken2025} inferred from using multiple \ce{CO2} bands to constrain the \ce{^12CO2}/\ce{^13CO2} ratio. Additionally, it is consistent with the range of values calculated towards low- and high-mass protostars \citep{Boogert2000, Brunken2024}.  Therefore, the high CO$_2$ ice abundance inferred from our model is compatible with the $^{13}$CO$_2$ absorption band depth considering an interstellar-like $^{12}$C/$^{13}$C ratio.

\section{Discussion}
\label{sec:discussion} 

In Section \ref{sec:results} we presented a model of IRAS 15398 that reproduces the JWST observed continuum along with solid-state silicate, H$_2$O, CO$_2$, and CO absorption features.  With this, we now aim to assess the extent to which radiative transfer effects influence the ice spectra observed towards protostars.

\subsection{Influence of ice on the silicate absorption}

\begin{figure}[h!]
    \centering
    \includegraphics[width=0.90\columnwidth]{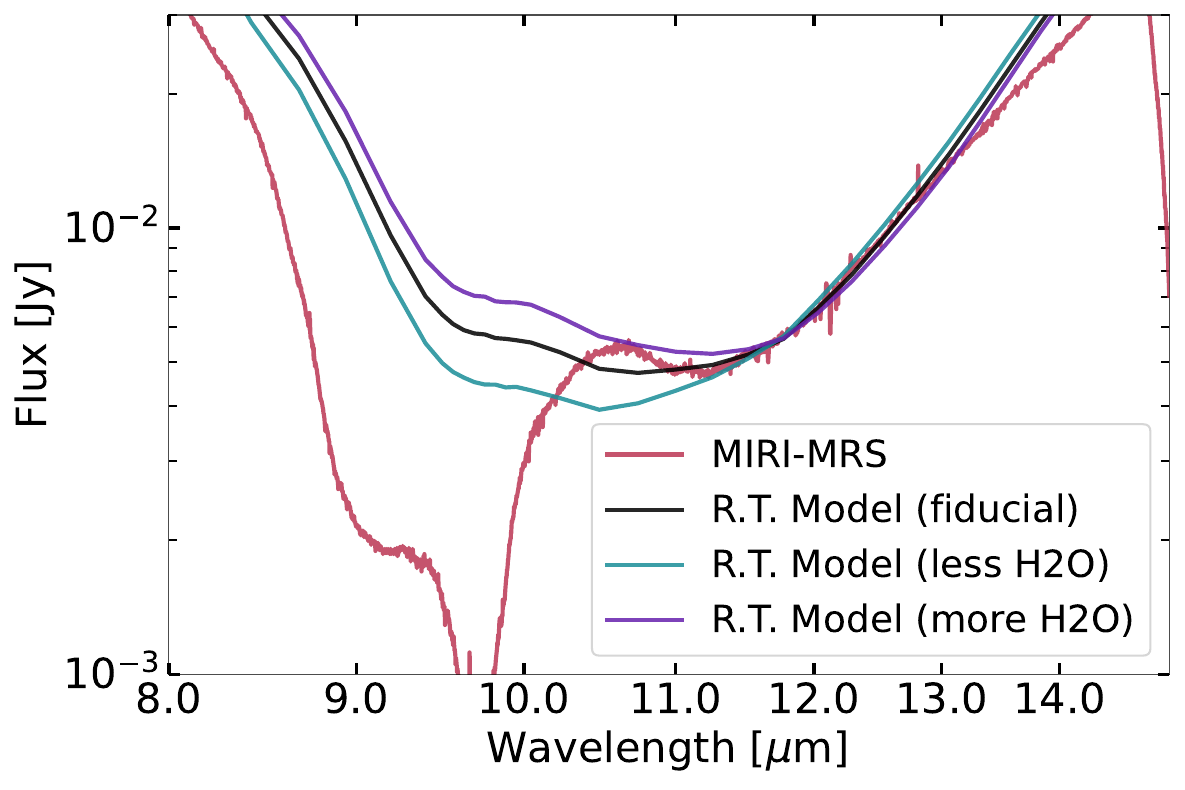}
    \caption{ Comparison of the fiducial model with models in which the \ce{H2O}a abundance is decreased to 1.9$\times10^{-4}$ (teal) or increased to 3.9$\times10^{-4}$ (purple). } 
    \label{fig:iceeffect}
\end{figure}

While silicate absorption features are commonly subtracted from the optical depth spectrum prior to fitting ice features, it is interesting to consider the influence of ice on the silicate absorption itself.  Previous work has noted that the presence of an ice mantle on dust grains mutes the absorption by silicates \citep[e.g.][]{Ossenkopf1994,Arabhavi2022}.  This occurs because light that could be absorbed by silicates can instead be absorbed or scattered by the ice mantles.  This can be seen directly in the opacities shown in Figure \ref{fig:opacities}, where the absorption at 10 $\mu$m of the dust-only population (Zone 5) is significantly reduced compared to Zones 1--4 where ices are present.  To illustrate the importance of this effect in our modeling, Figure \ref{fig:iceeffect} shows our preferred model of IRAS 15398 compared with models in which the amorphous water abundance is increased and decreased to  3.9$\times10^{-4}$ and 1.9$\times10^{-4}$ respectively, from the fiducial value of 2.9$\times10^{-4}$. 

As the ice abundance is increased in the model, the silicate absorption feature becomes shallower.  Because the 10 $\mu$m silicate absorption and 11.5 $\mu$m librational \ce{H2O} band overlap, adding more \ce{H2O} onto the mantle will counter-intuitively decrease the total optical depth in the range of 9--12 $\mu$m.  In our model tuning, we found that we had to carefully balance the ice and dust properties and source physical structure in order to maintain consistency with this part of the spectrum. Efforts to isolate ice absorption features by subtracting a silicate profile should consider the importance of self-consistently treating the silicate and ice opacities, and using all available ice and silicate absorption features in the near- and mid-IR range to break degeneracies when possible.

\subsection{Continuum identification and optical depth determination}
\label{sec:tau_compare}
\begin{figure*}

    \centering
    \includegraphics[width=0.65\paperwidth]{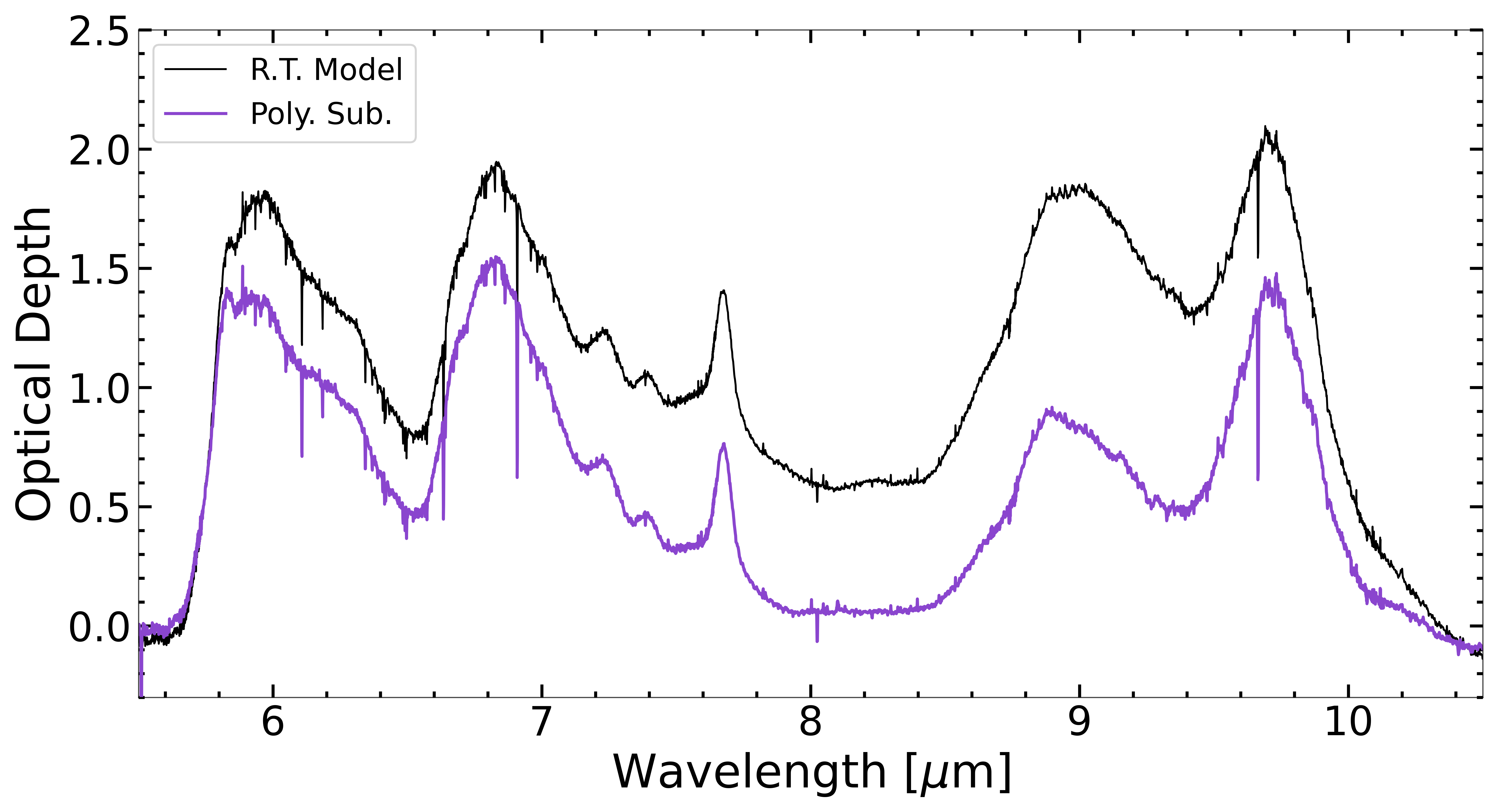}
    \caption{Optical depth spectra made from a polynomial continuum model of IRAS 15398 \citep[purple;][]{Kim2025}, and by using the radiative transfer model (black). } 
    \label{fig:opticaldepthspectra}
\end{figure*}

Identifying the continuum is one of the most uncertain steps in quantifying low-abundance ice features in protostars \citep{Rocha2025,Gross2025}. The fit may be influenced by accumulated ice features, especially in the crowded 6-8 \micron\ region. In addition, the usual method assumes that all the continuum arises behind the ice absorption; emission mixed with or in front of icy dust could veil the observed ice features. Spectral resolution effects will also impact the ability to accurately fit a polynomial continuum. Finally, different ice features are likely to arise in different regions along the line of sight, which might affect inferred ratios of ice abundances.

 Here, we aim to understand how the commonly-used approach of fitting a polynomial continuum model compares to our continuum model, which is constrained by a broad-band fit across the near-/mid-IR along with the SED.  We are particularly interested in the impact of the continuum choice on the calculated optical depths in the $\sim$6--10$\mu$m range, which contains many overlapping features of trace ice components.  Optical depth spectra are shown in Figure \ref{fig:opticaldepthspectra}, calculated using Equation \ref{odequation} with either our radiative transfer model or a polynomial model \citep{Kim2025} as the continuum flux. The polynomial model accounts for pyroxene (Mg$_{0.7}$Fe$_{0.3}$SiO$_{3}$) and olivine (\ce{MgFeSiO4}) silicate components, as well as crystalline and amorphous water ice.  Thus, absorption due to silicates and water ice has been removed in the optical depth spectra of both the radiative transfer and polynomial models, and the two can be directly compared in the wavelength range shown in Figure \ref{fig:opticaldepthspectra}.  A comparison of both model fits to the observed flux spectrum is shown in Appendix \ref{appendixRTvsPoly}.

There is a considerable difference in the optical depth spectra calculated with the two different continuum models.  While all of the same absorption features are present in both cases, their optical depths are generally higher in the radiative transfer version.  Moreover, the feature shapes are in some cases different, such as around 6, 8.5, 9, and 10.5~$\mu$m. The higher optical depth in the radiative transfer spectrum from 5.5--7.5 $\mu$m could imply more organic material, such as the ``C5'' component in \cite{Boogert2008,Boogert2015}. More of this absorption is filled in by water ice in the polynomial fit. The difference between the radiative transfer and polynomial spectra is highest around $\sim$9~$\mu$m, where more of the optical depth is attributed to silicate absorption in the polynomial spectrum. Pyroxene can contribute additional optical depth from $\sim$8.5--10 $\mu$m, which could explain some of the difference in this range between the polynomial and radiative transfer optical depth spectra. Overall, both the absolute ice column densities and the column density ratios will strongly depend on whether a polynomial or radiative transfer continuum is adopted.  The radiative transfer continuum is both physically informed and constrained by other observations of the source, but still may not perfectly describe the true continuum of the source.  Rather, we advise caution in the interpretation of ice abundances extracted from weak features, especially in the $\sim$8--10~$\mu$m range.

\subsection{Origin of ice absorption within the envelope}
\label{subsec:ice_origins}
Our radiative transfer framework provides an opportunity to evaluate what region of the protostellar envelope is actually being traced by the observed ice absorption features.  Following the `contribution function' (hereafter CF) method described in \cite{Sturm2024}, we calculated the fraction of each ice absorption feature that originates from different regions in the envelope model. The CF models were performed by dividing the model grid into spatial zones, and removing ices from all but one zone at a time.  Synthetic spectra were ray traced to determine the amount of absorption contributed by each zone individually across the whole model. The optical depth for each zone is calculated using Equation \ref{odequation} and expressed as a percentage of the optical depth for the model with ice in all zones (i.e.~Figure \ref{fig:spectrum}). 

\begin{figure}[]
    \centering
    \includegraphics[width=\linewidth]{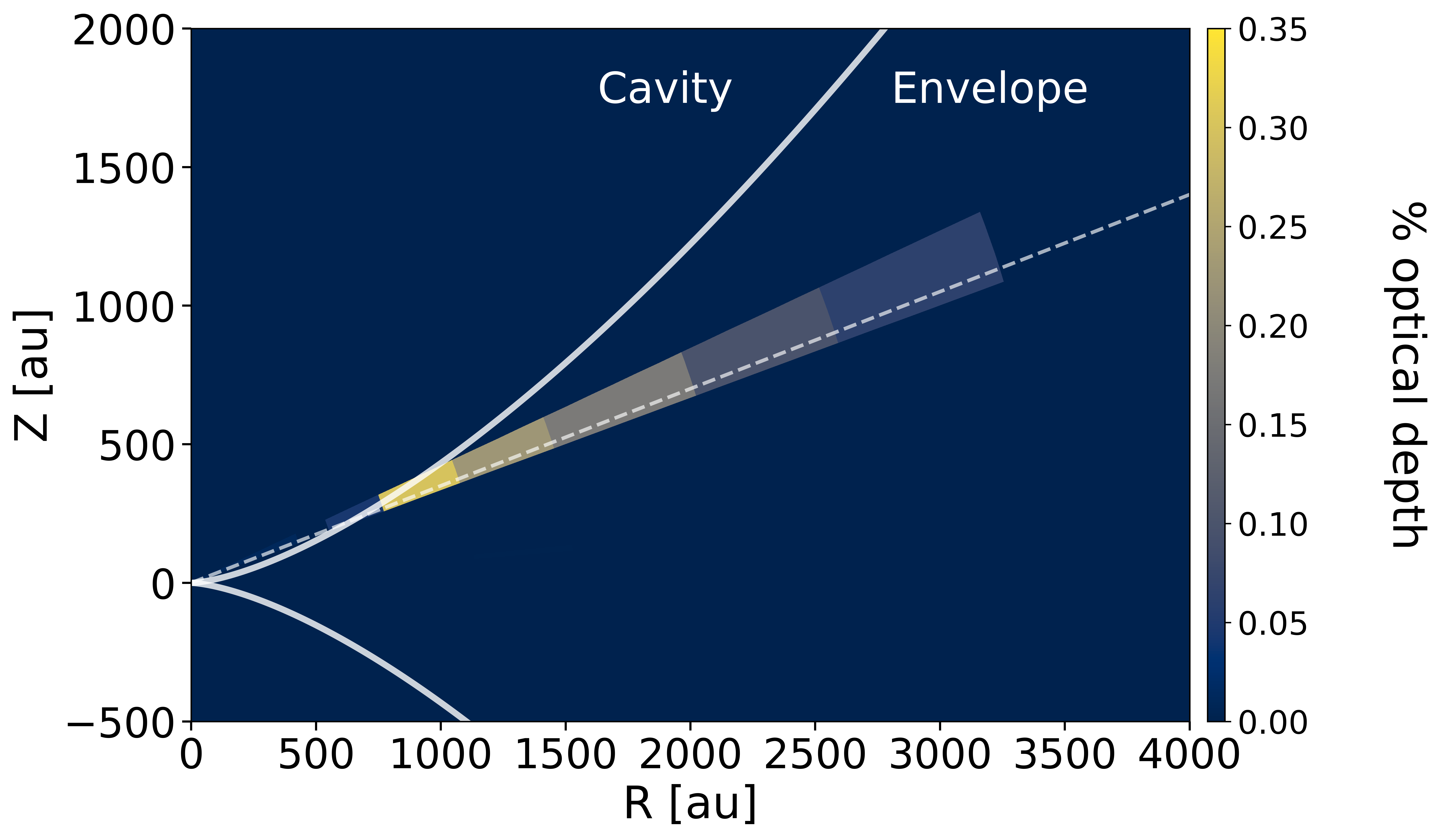}
    \caption{Contribution function plot for the 15 $\mu$m \ce{CO2} absorption band, showing the fraction of the total optical depth contributed by different regions in the model. The cavity/envelope boundary is shown by the white contour, and the white dashed line shows the line of sight based on the viewing inclination angle of 71$^{\circ}$ for IRAS 15398. }
     \label{fig:cbfmir}
\end{figure}

Figure \ref{fig:cbfmir} shows the CF map for the \ce{CO2} bending mode at 15 $\mu$m.  The sum of optical depths in the CF map is 92\% of the optical depth in the full model, confirming that this is a valid approach to assessing absorption contributions. We also ran a CF analysis for the near-IR CO$_2$ (4.7 $\mu$m) and H$_2$O (3 $\mu$m) bands, and found no discernible difference compared to Figure \ref{fig:cbfmir}, indicating that all 3 absorption features are tracing a similar region of the inner envelope.

\begin{figure}
    \centering
    \includegraphics[width=0.90\columnwidth]{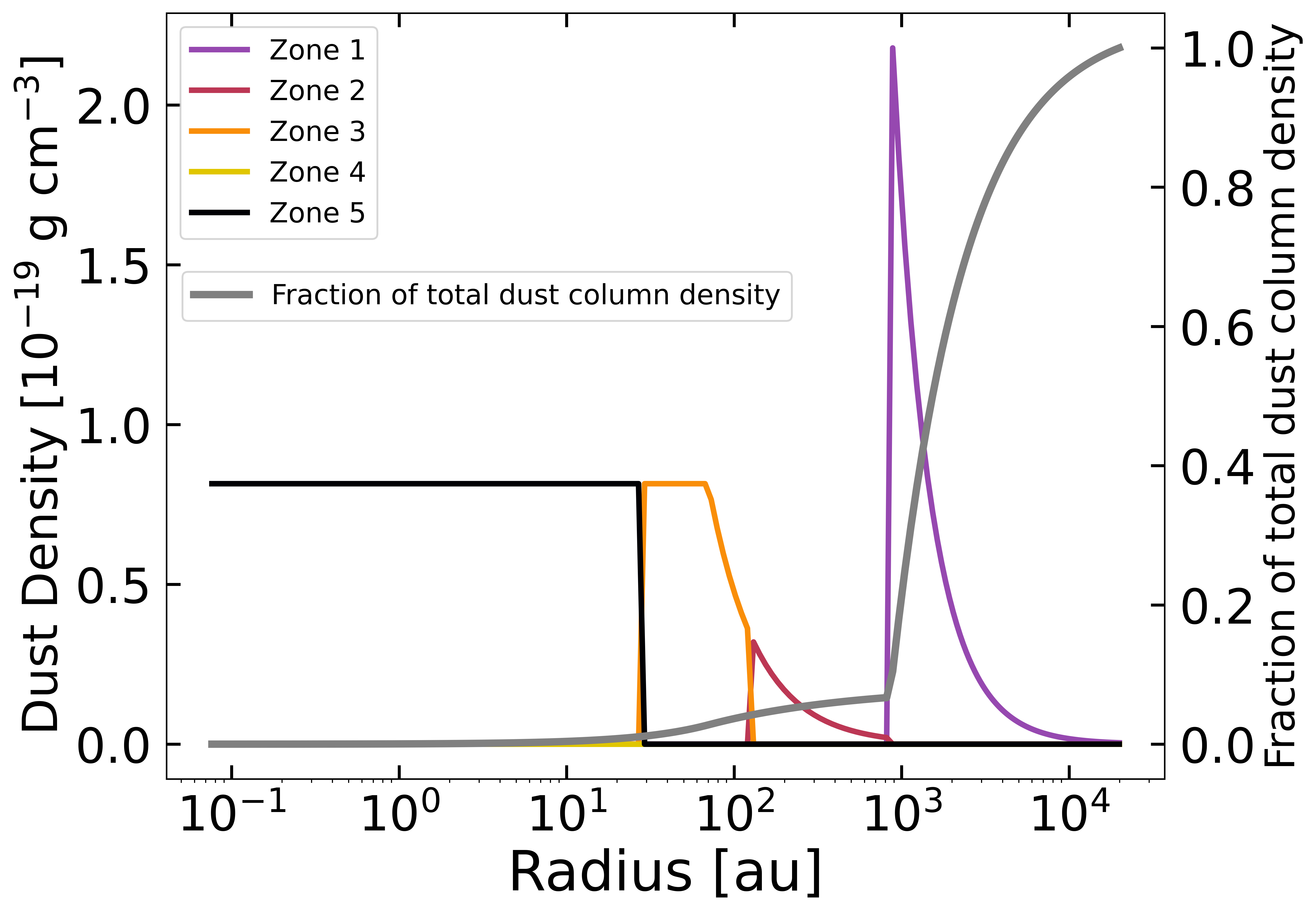}
    \caption{Dust density (g cm$^{-3}$) for each dust + ice population along the line of sight. The cumulative dust density for all dust + ice populations along the line of sight is overlaid in gray. } 
    \label{fig:losdensitycumulative}
\end{figure}

The absorption occurs entirely along the viewing line of sight towards the central star, peaking around a radius of 1000 au and with smaller contributions extending out to $\sim$3000 au.  The highest absorption contribution coincides with the transition from the cavity to the envelope, indicated by the white contour.  As expected, there is almost no ice absorption within the cavity region, and it is noteworthy that the majority of the absorption is located at small distances from the protostar rather than in the outer envelope. The ubiquitous, strong 15 $\mu$m CO$_2$ absorption in \cite{Furlan2008} was attributed to a \ce{CO2} origin in the cold outer envelope, which does not agree with the results of our CF analysis. To better understand this, Figure \ref{fig:losdensitycumulative} shows the density of solids encountered along the line of sight, with local densities shown in colors and the cumulative column density shown in gray.
  
At the smallest radii, the line of sight passes through the constant-density region of the cavity, which is ice-free (Zone 5) from the dust sublimation radius ($\sim$0.1 au) to 30 au. Beyond this, water ice is present on the solids (Zone 3 \& 4).  At 70 au, the density begins to decrease as the power-law part of the cavity begins, with CO$_2$ ice (Zone 2) freezing out just past 100 au. Finally, at 845 au, the line of sight enters the envelope, indicated by a sharp rise in density and the appearance of CO ice (Zone 1), followed by a decrease in the density throughout the rest of the envelope. 

Between the infall radius of 845 au and the outer radius of the model (20,000 au), the density is described by a power-law: 
\begin{equation}
n(r) = n_{f}(r/r_{f})^{-\alpha}
\end{equation}
where n$_{f}$ is the fiducial density at a fiducial radius r$_{f}$. 
For $\alpha > 1$, the column density can be approximated as:
\begin{equation}
N = \frac{n_f r_{f}^{\alpha}}{\alpha-1} r_{i}^{1-\alpha} (1-(r_i/r_o)^{\alpha -1}) \simeq \frac{n_f r_{i}}{\alpha-1} 
\end{equation}
where $r_{i}$ and $r_{o}$ are the inner and outer radius, respectively, and the approximation applies for $r_{o} >> r_{i}$ and we take $r_f = r_i$. The fact that the outer radius disappears in the approximation shows that the column density and thus the optical depth are dominated by the innermost regions for a power law with $\alpha > 1$.
For our TSC model ($\alpha = 2.0$), the column density depends strongly on the infall radius and hardly at all on the outer radius. The cumulative density in Figure \ref{fig:losdensitycumulative} indicates that half the total ice column is reached by 1530 au, only 8\% of the outer radius, so the absorption spectrum is actually determined by the density in the innermost regions where the ice exists. For our model of this source, that is where the line of sight enters the envelope. The outer envelope contains a lot of ice, but the absorption spectrum is not sensitive to it. This result is a feature of the envelope density structure and does not depend on the nature of the cavity model.

\subsubsection{Validity of pencil-beam absorption}
The fact that the the 3 $\mu$m H$_2$O, 4.7 $\mu$m CO$_2$, and 15 $\mu$m CO$_2$ absorption features all originate from the same region of the envelope supports the idea that the continuum forms at much smaller radii than the ices being traced.  Otherwise, we would expect the onset of absorption to vary with wavelength.  Overall, it appears that treating protostellar ice absorption as pencil-beam absorption against a point-source continuum is valid. Indeed, this is expected because the temperature needed to emit strongly at even 15 \micron\ is much higher than the ice sublimation temperature.  However, this may be an issue for observations of ice features in the far-infrared.  Indeed, in disks the contribution function of ice features has been shown to depend strongly on the inclination angle \citep{McClure2015}.

\subsubsection{Line-of-sight column density vs.~abundance ratios}
\begin{figure}
    \centering
    \includegraphics[width=0.90\columnwidth]{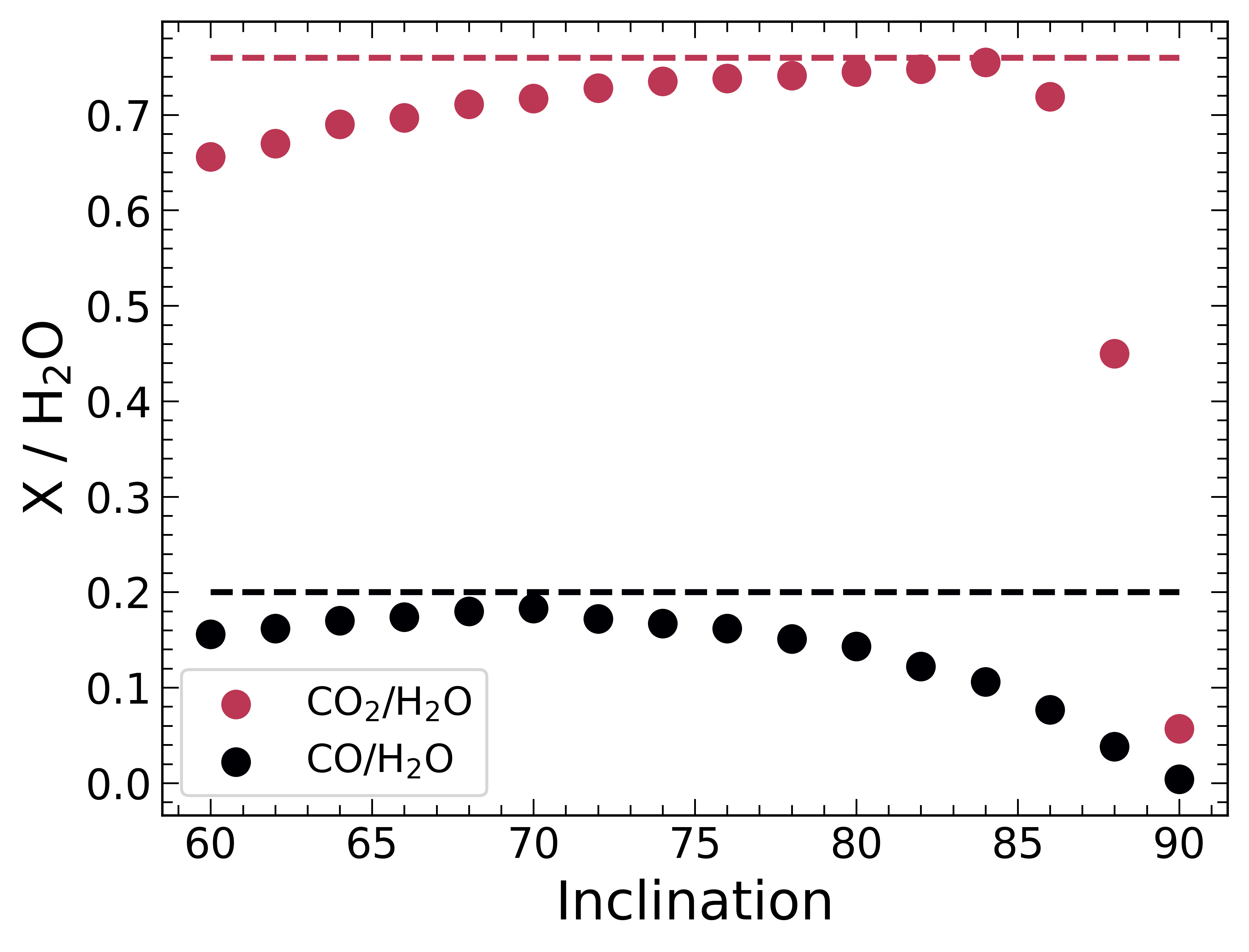}
    \caption{Column density ratios vs. inclination for \ce{CO2}/\ce{H2O} and CO/\ce{H2O} based on our fiducial radiative transfer model. The modeled abundance ratios of 0.76 for \ce{CO2} and 0.20 for CO are shown as horizontal lines for reference.} 
    \label{fig:cdratios}
\end{figure}

This analysis can also explain why the CO/H$_2$O and CO$_2$/H$_2$O column density ratios calculated along the viewing line of sight are nearly identical to the inputted abundance ratios, as described in Section \ref{subsec:ice_columns}.  Although the line of sight passes through regions where H$_2$O is present without CO$_2$ or CO, in our model these zones occur entirely within the cavity and contribute minimally to the observed absorption. However, this will not always be the case, depending on the combination of viewing angle, cavity geometry, and temperature structure. For illustration, Figure \ref{fig:cdratios} shows the line-of-sight column densities calculated by our model if IRAS 15398 was observed at different viewing inclinations.  At our fiducial model's inclination of 71$^{\circ}$, the calculated \ce{CO2}/\ce{H2O} and CO/\ce{H2O} column density ratios of 0.72 and 0.18 are close to the input abundance ratios of 0.76 and 0.2.  However, the ratios can deviate significantly at higher or lower inclinations, especially near 90$^\circ$ when the line of sight does not pass through the cavity and a significant fraction of the total absorption originates from the zones where H$_2$O is frozen without CO$_2$ or CO.  Therefore, while this dilution effect is not very important for IRAS 15398, it may be quite important for other sources; in these cases the line-of-sight column densities will underestimate the true ice abundance ratios where all molecules are frozen out.

\subsection{The high CO$_2$ ice abundance}

As described in Section \ref{subsec:ice_columns}, our model persistently required a high CO$_2$/H$_2$O abundance ratio to explain the depth of the observed 15 $\mu$m CO$_2$ band.  Our fiducial model's preferred CO$_2$/H$_2$O abundance of 76\% is supported by the fact that it reproduces the depth of the $^{13}$CO$_2$ stretching mode without the need to invoke extreme carbon fractionation.  Source-specific radiative transfer modeling of additional protostars is needed to understand whether such a high CO$_2$ abundance is common, and whether some observational effect not considered here could be at play.  Note that one prior work, \citet{Furlan2008}, carried out radiative transfer modeling of 28 Class I protostars based on their observed IRS spectra, which included fitting a CO$_2$ ice abundance.  However, they kept the H$_2$O ice abundance fixed in all models, and therefore the CO$_2$/H$_2$O ratios derived from that work are not well constrained.  

If the high CO$_2$/H$_2$O inferred for IRAS 15398 is confirmed, it does not necessarily mean that the CO$_2$/H$_2$O ice ratio is globally enhanced throughout the entire envelope.  Indeed, our CF analysis shows that the observations primarily reflect absorption at the cavity-envelope interface ($\sim$1000-2000 au) rather than the envelope-averaged ice composition along the line of sight.  The cavity-envelope interface in particular could be a natural site for enhancement of \ce{CO2} due to steep temperature and density gradients, as well as higher exposure to irradiation than the shielded outer envelope.  These environmental conditions can increase the availability of OH radicals from \ce{H2O} photodissociation, promoting grain-surface conversion of CO into \ce{CO2}. Enhanced \ce{CO2} production driven by photochemical reactions has been suggested from Spitzer observations of low-mass YSOs \citep{Cook2011}, as well as laboratory studies of CO and non-energetic OH radicals \citep{Oba2010}.  The double-peaked profile of the 15~$\mu$m CO$_2$ band in IRAS 15398 also indicates that the ice has undergone thermal processing \citep[e.g.][]{Gerakines1999,Pontoppidan2008}, which could similarly reflect local alteration of the ice composition.

We also note that high \ce{CO2}/\ce{H2O} ratios have been observed in multiple comets, including the solar system comet C/2016 R2 and the extrasolar comet 3I/ATLAS, which had coma \ce{CO2}/\ce{H2O} mixing ratios of 29 and 7.6, respectively \citep{Cordiner2025}. While multiple scenarios for planetesimal formation and evolution could lead to a high coma \ce{CO2}/\ce{H2O} mixing ratio, these results show that the ratio of $\sim$0.76 inferred for IRAS 15938 is well within the range that is possible in primitive icy materials. 

\section{Conclusions}

We have developed a new radiative transfer framework to simulate observations of protostellar ice absorption.  We present a model tailored to the IRAS 15398 protostar observed by the JWST CORINOS program, for which the source physical structure, dust properties, and ice abundances have been tuned to reproduce the observed SED, IR continuum, and absorption features from silicates and H$_2$O, CO$_2$, and CO ice.  Our main conclusions are as follows: 
\begin{itemize}[noitemsep,leftmargin=*]
    \item Using the radiative transfer model, we find a H$_2$O column density and CO/H$_2$O column density ratio similar to those found with empirical methods.  However, our model requires a CO$_2$/H$_2$O column density ratio around 70\%, which is twice as high as the values previously found for this source.  
    \item The high CO$_2$ abundance is supported by the depth of the $^{13}$CO$_2$ stretching feature, which is reproduced with a $^{12}$CO$_2$/$^{13}$CO$_2$ abundance ratio of 118.
    \item The continuum determined from our radiative transfer model differs considerably from a simple polynomial fit, and produces different optical depths for the absorption features in the 6--10 $\mu$m range.  Due to uncertainties in the continuum, caution must be taken in interpreting the shape and depth of these features.
    \item The presence of an ice mantle mutes the 10 $\mu$m silicate absorption feature: counterintuitively, higher H$_2$O ice abundances decrease the optical depth at the 11.5 $\mu$m H$_2$O libration mode.  This effect should be considered in empirical approaches which subtract off a silicate profile prior to fitting the ice features.
    \item Most of the ice absorption in the model originates along the viewing line of sight at a radius of 1000-2000 au, peaking where the line of sight transitions from the cavity to the envelope.  The absorption spectrum for this source is not sensitive to the large column of ice in the outer envelope.  This is explained by the fact that the absorption predominantly traces the inner-most region where the ice density is described by a power-law distribution, which for this source occurs where the line of sight enters the envelope.  
    \item For our preferred model of IRAS 15398, the line-of-sight column density ratios are nearly the same as the ice abundance ratios in the region where all ices are frozen out.  This is because differential freeze-out along the line of sight only occurs within the cavity, which contributes minimally to the total absorption.  However, for different viewing geometries or source structures, the CO$_2$/H$_2$O and CO/H$_2$O column density ratios may significantly underestimate the true ice abundance ratios.
    
\end{itemize}

\begin{acknowledgments}
\end{acknowledgments}

This work is based on observations made with the NASA/ESA/CSA James Webb Space Telescope. The data were obtained from the Mikulski Archive for Space Telescopes at the Space Telescope Science Institute, which is operated by the Association of Universities for Research in Astronomy, Inc., under NASA contracts NAS 5–26555 and NAS 5-03127. The JWST data presented in this article were obtained from the Mikulski Archive for Space Telescopes (MAST) at the Space Telescope Science Institute. The specific observations analyzed can be accessed via \dataset[doi: 10.17909/929r-rh82]{https://doi.org/10.17909/929r-rh82}.

Support for Program number 2151 was provided by NASA through a grant from the Space Telescope Science Institute, which is operated by the Association of Universities for Research in Astronomy, Inc., under NASA contract NAS 5-03127. 
Y.-L.Y. and N.S acknowledges support from Grant-in-Aid from the Ministry of Education, Culture, Sports, Science, and Technology of Japan (20H05845 and 25H00676), and a pioneering project in RIKEN (Evolution of Matter in the Universe). A part of this research was carried out at the Jet Propulsion Laboratory, California Institute of Technology, under a contract with the National Aeronautics and Space Administration (80NM0018D0004).

\software{RADMC-3D \citep{Dullemond2012}, OpTool \citep{Dominik2021}, Matplotlib \citep{Hunter2007}, NumPy \citep{harris2020}, Pandas \citep{pandas2020}
          }

\appendix
\section{Ice and dust parameters}
\label{appendixconstants}

The molecular constants used in the dust + ice modeling, including binding energies, specific densities, and molar masses, are reported in Table \ref{table:constants}.  The binding energies for \ce{H2O}a, \ce{H2O}c, \ce{CO2}, and CO were chosen as the recommended binding energies from \citet{Minissale2022}. For \ce{CO2} and CO, values are for \ce{CO2} and CO on amorphous solid water.

\begin{table}[h!]
\centering
\caption{Constants used in the dust + ice model}
\begin{tabular}{c c c c} 
\hline\hline
 Species & E$_{i}$ (K) & Specific Density (g cm$^{-3}$) & Molar mass \\ [0.5ex] 
 \hline
 \ce{H2O}a    &  5640$^{a}$ &  0.87$^{b}$  &   18  \\ 
 \ce{H2O}c    &  5803$^{a}$  &  0.92$^{b}$  &   18  \\
 \ce{CO2}     &  3196$^{a}$  &  1.67$^{c}$  &   44  \\
 CO           &  1390$^{a}$   &  0.80$^{d}$  &   28  \\
 \ce{MgFeSiO4} & -          & 3.3$^{e}$      & -    \\
 Carbon (amorphous)    &     -     & 1.8$^{f}$     & -     \\
 Carbon (graphite)    &     -     & 2.16$^{g}$     & -     \\
 \hline
\end{tabular}
\label{table:constants}
\tablenotetext{}{$(a)$ \cite{Minissale2022}, 
$(b)$ \cite{Dohnalek2003}, 
$(c)$ \cite{Loeffler2016}, 
$(d)$ \cite{Bouilloud2015}, 
$(e)$ \cite{Draine2003astrosil}, 
$(f)$ \cite{Zubko1996}, 
$(g)$ \cite{Draine2003graph}}
\label{constants}
\end{table}

\FloatBarrier
\section{Photometry data}
\label{appendixSED}
Table \ref{tab:photometry} lists the photometry data for IRAS 15398 used to construct the SED shown in Figure \ref{fig:sed}.  Near-IR data points are taken from the 2MASS \citep{Skrutskie2006} and Spitzer \citep{Evans2009} archives, and mid-IR points from the CORINOS MIRI/MRS observations.  Far-IR data is taken from the Herschel archive, Herschel Gould Belt Survey \citep{Andre2010}, and \cite{Green2016}, sub-millimeter data from SCUBA via \cite{Shirley2000}, and the millimeter point from \cite{nurnberger1997}.

\begin{table}[h!]
\centering
\caption{Photometric Data \label{tab:photometry}}
\begin{tabular}{c c c c c} 
\hline\hline
 Wavelength ($\mu$m) & Flux (Jy) & Flux error (Jy) & Aperture (arcsec.) & Reference \\ [0.5ex] 
 \hline
 1.662      & 3.21$\times10^{-4}$  & 3.21$\times10^{-4}$  & 20.0  &   2MASS  \\ 
 2.159     & 3.95$\times10^{-4}$  & 3.95$\times10^{-4}$ &  20.0   & 2MASS \\
 3.6        & 0.00106  & 1.1$\times10^{-4}$ &  3.0      &       IRAC    \\
 4.5       & 0.00953  & 9.5$\times10^{-4}$ &  3.0      &              IRAC \\
 5.6       & 0.016  & 0.0016  &  1.74      &       JWST MIRI/MRS \\
 8.0       & 0.071  & 0.0071 &  2.48        &      JWST MIRI/MRS \\
 22.6      & 0.785  & 0.0785 &  6.82        &      JWST MIRI/MRS \\
 70.0      & 20.25  & 2 &  32               &        Herschel Gould Belt Survey \\
 100.0      & 43.55  & 4.36  &  32          &      Herschel Science Archive  \\
 160.0     & 50.316  & 5.03 &  32   &          Herschel Gould Belt Survey \\
 250.0     & 44.2  & 10.8 &  39.5     &   \cite{Green2016}      \\
 350.0     & 27.5  & 8.39 &  43.1      &  \cite{Green2016}          \\
 450.0      & 19.6  &  2.7 &  40      &       SCUBA  \\
 450.0     & 25.9  & 3.7 &   120    &         SCUBA   \\
 500.0      & 14.5 & 6.31 & 50.7    &          \cite{Green2016} \\
 850.0      & 2.63  & 0.25  &  40      &        SCUBA    \\
 850.0      & 4.23  & 0.24 &  120      &         SCUBA   \\
 1300.0     & 0.365  & 0.01  & 24      &     \cite{nurnberger1997}        \\
 \hline
\end{tabular}
\label{sedtable}
\end{table}

\FloatBarrier
\section{Source age determination}
\label{appendixage}
\begin{figure}[ht!]
\includegraphics[width=0.45\textwidth]{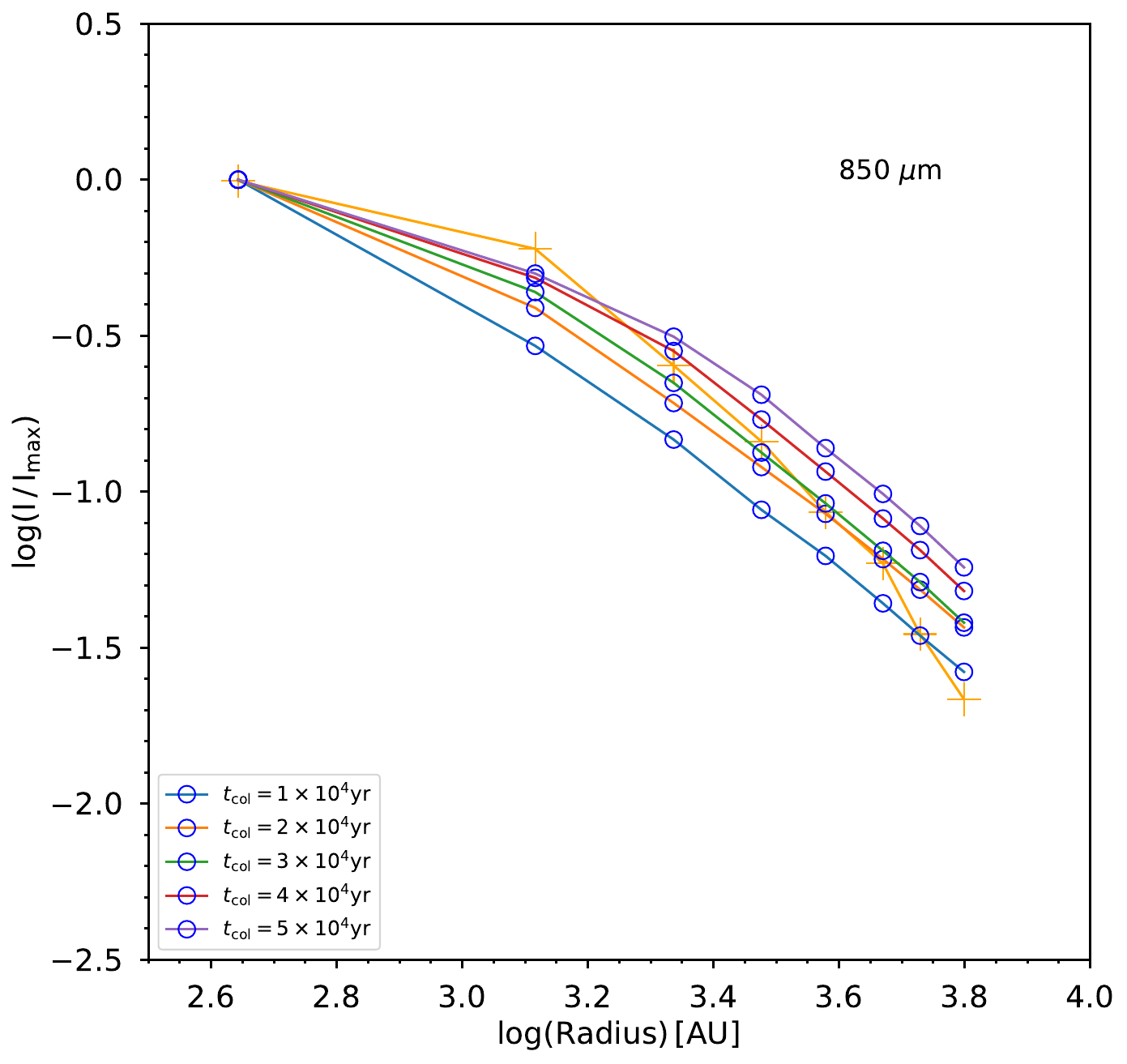}
\caption{
The radial profile of the observations at 850 $\mu$m, digitized from Figure 10 of \citet{Shirley2000}, is shown in yellow. The other curves are predictions from models of differing ages.
}
\label{radprof}
\end{figure}

The age of the system in the TSC model is the time since the initial point source formed. As the wave of infall moves outward at the effective sound speed, the distribution of matter changes. The initial density distribution ($n \propto r^{-2}$) becomes flatter, approaching $n \propto r^{-1.5}$ inside the infall radius. This changes both the SED and the radial profile of submillimeter emission, with the radial profile being the most diagnostic of age \citep{Shirley2002}. In preliminary modeling, different ages were tried, as shown in Figure \ref{radprof}. None of the models fit all the points well, but an age of $3 \times 10^4$ yr clearly fits better over most of the range.

\FloatBarrier
\section{Dust temperature structure}
\label{appendixdusttemp}

\begin{figure*}[h!]
    \centering
    \includegraphics[scale=0.35]{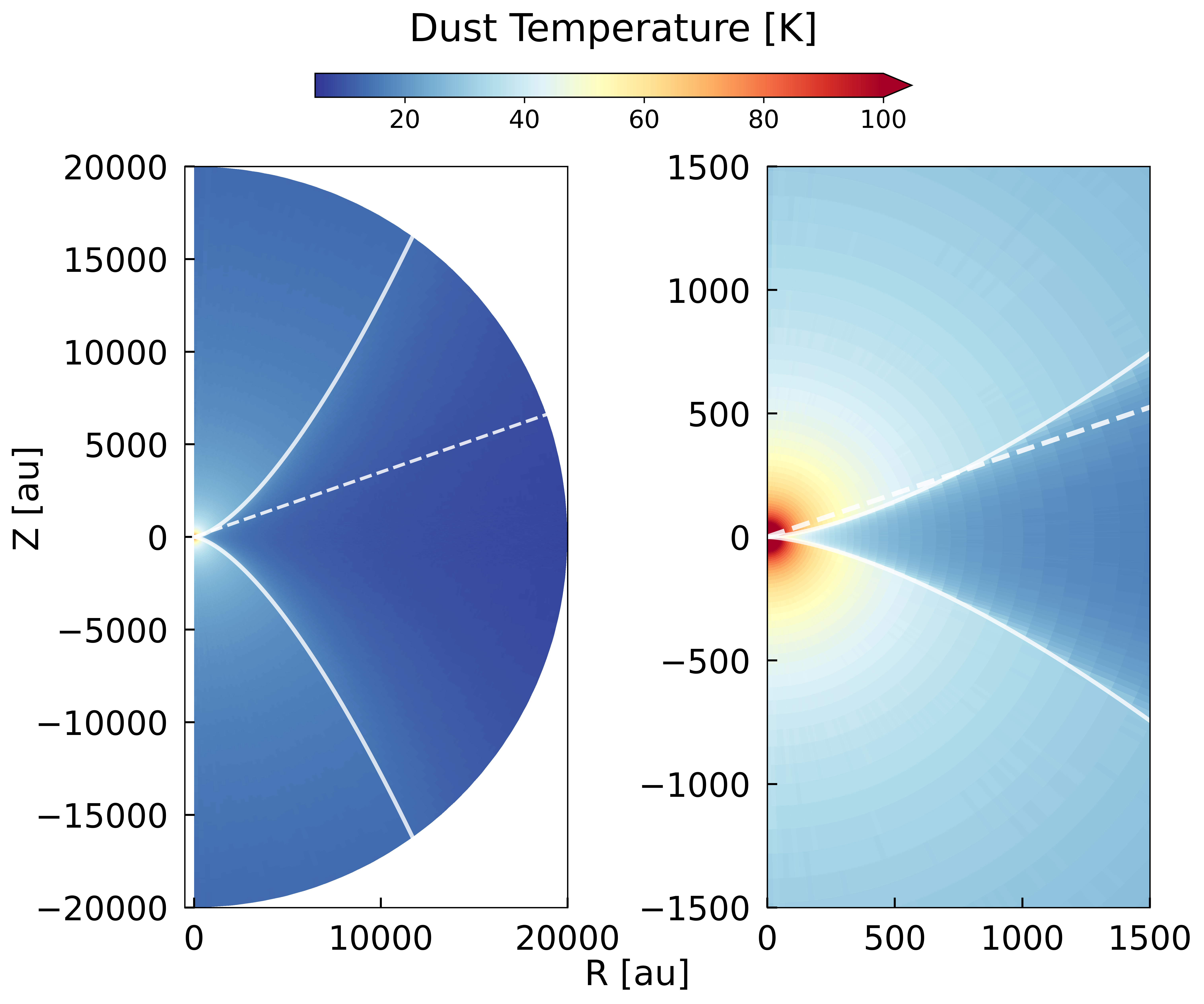}
    \caption{Dust temperature structure in our preferred model of IRAS 15398. The white dashed line visualizes the pencil-beam line of sight based on the inclination of IRAS 15398 (71$^{\circ}$), and the white contour displays the cavity-envelope boundary in the model.  The right panel of the figure is zoomed in around the inner 1500 au of the protostar.} 
    \label{fig:2ddusttemp}
\end{figure*}

Figure \ref{fig:2ddusttemp} shows the temperature structure calculated from the thermal Monte Carlo calculation for our preferred model of IRAS 15398, analogous to the density structure shown in Figure \ref{fig:2ddustdensity}.

\FloatBarrier
\section{Near-IR ice features}
\begin{figure}[h!]
    \centering
    \includegraphics[scale=0.6]{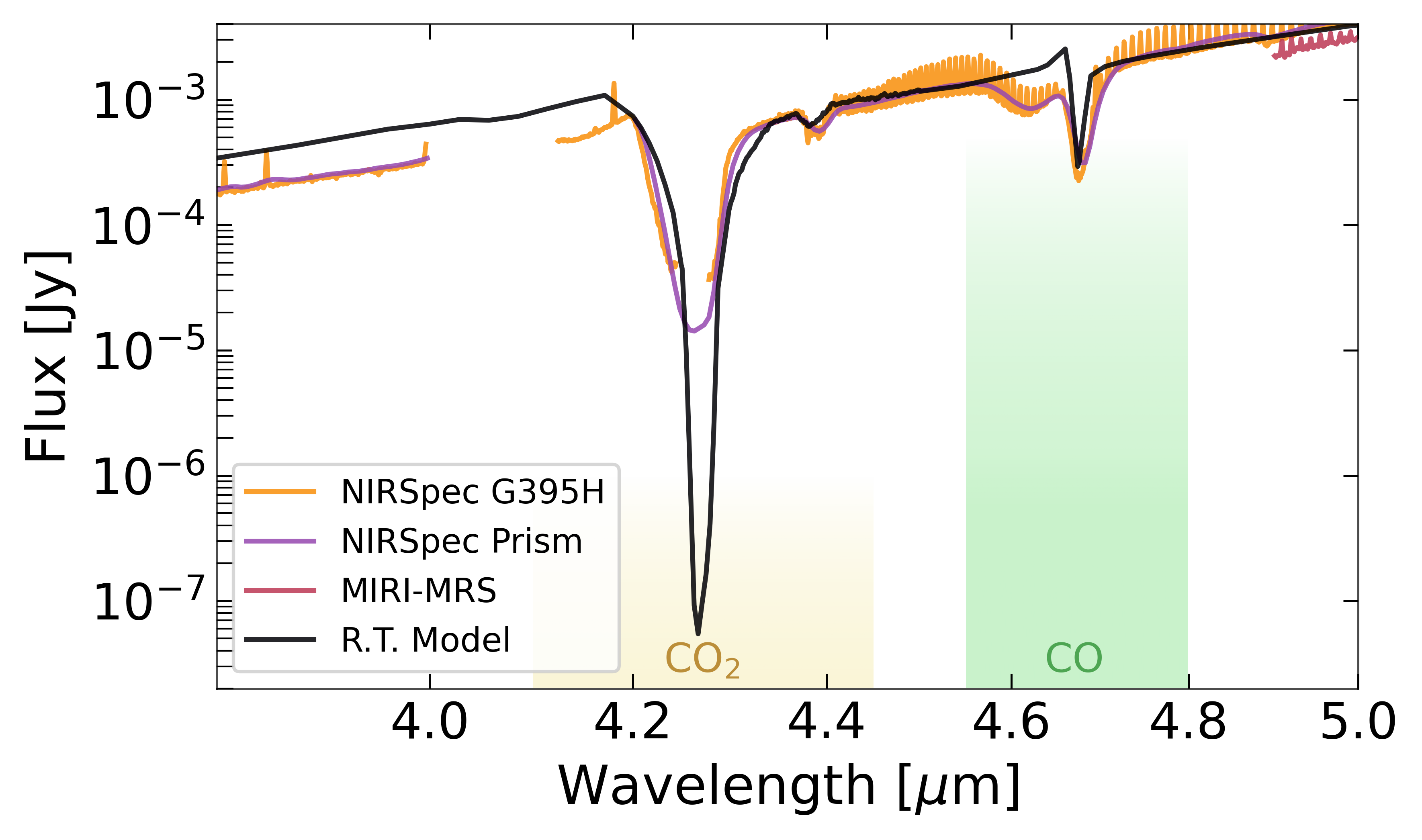}
    \caption{A zoomed in version of Figure \ref{fig:spectrum} around the NIR \ce{CO2} and CO absorption bands. Shown are the NIRSpec spectrum (Program ID 1854, PI: M. McClure), NIRSpec Prism spectrum (Program ID 6161, PI: K. Slavicinska), and MIRI-MRS spectrum (Program ID 2151, PI: Y.-L. Yang) with the radiative transfer model spectra overlaid in black. Data points which hit the noise floor from the Prism spectrum were removed.} 
    \label{fig:co2zoom}
\end{figure}

To better visualize the \ce{CO2} and CO absorption features shown in Figure \ref{fig:spectrum}, Figure \ref{fig:co2zoom} presents a zoomed in version from 3.8 -- 5.0 $\mu$m. The model over-predicts the depth of the CO$_2$ stretching mode, but based on the good fit to $^{13}$CO$_2$ stretching mode and the CO$_2$ bending mode, this likely reflects saturation in the observations (Section \ref{subsec:ice_columns}).

\label{appendixRTzoom}
\FloatBarrier

\section{Radiative transfer vs.~polynomial continuum models}
\label{appendixRTvsPoly}

\begin{figure}[h!]
    \centering
    \includegraphics[width=0.90\columnwidth]{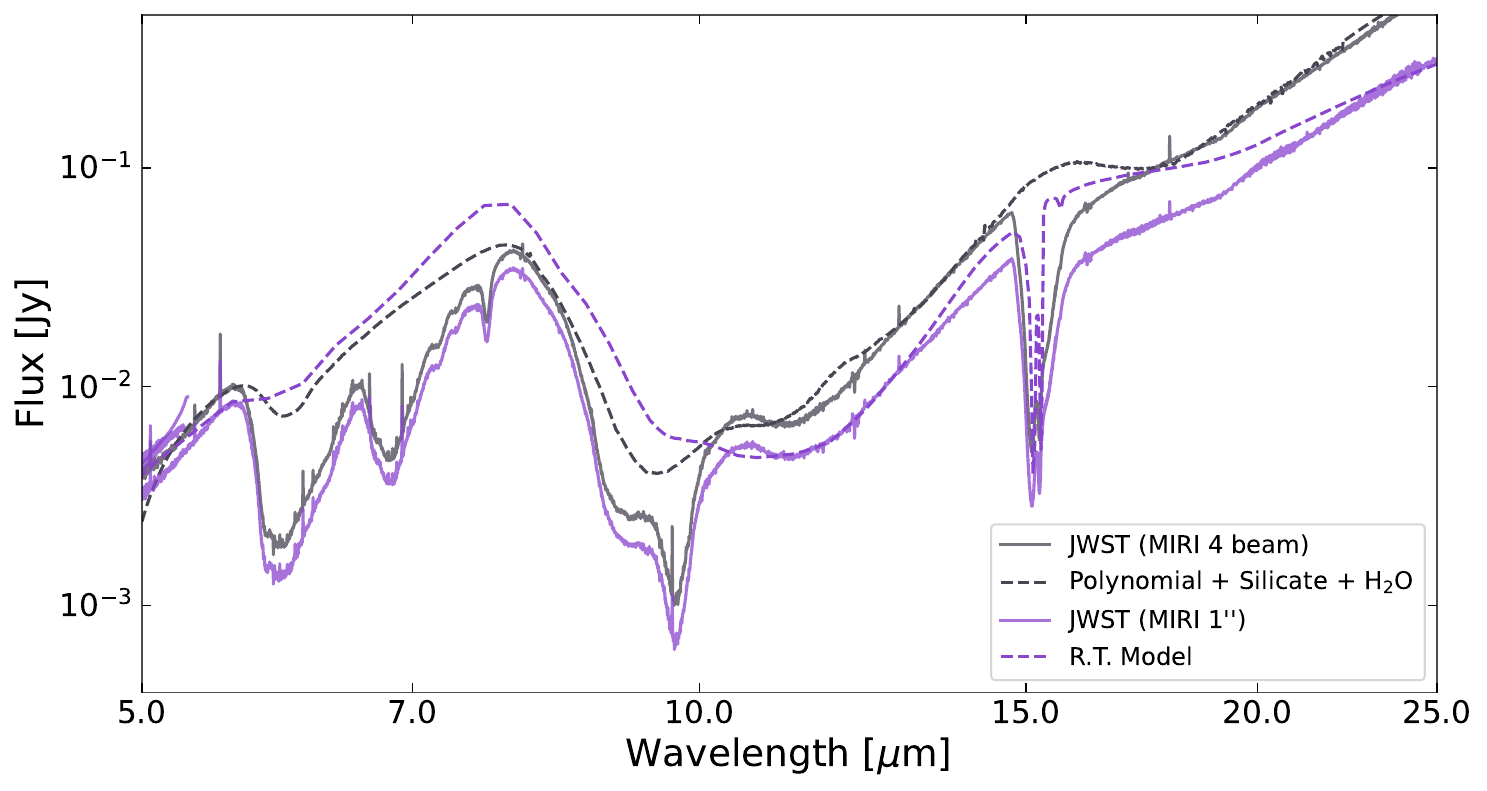}
    \caption{A comparison of the radiative transfer model (purple dashed) with the polynomial continuum model (black dashed) used to make the optical depth spectra shown in Figure \ref{fig:opticaldepthspectra}. The $1''$ diameter aperture extracted spectrum was used for this analysis (purple solid), while the 4 beam extracted spectrum was used in \cite{Kim2025}. }
    \label{fig:rtvspoly}
\end{figure}

Figure \ref{fig:rtvspoly} shows the JWST data compared to our radiative transfer model and the polynomial + silicate + water continuum model from \cite{Kim2025}.  Note that the polynomial continuum model was fit to the JWST MIRI/MRS spectrum extracted with a 4-beam aperture, while the radiative transfer model in this work was fit to the spectrum extracted with a fixed $1''$ diameter aperture.

\FloatBarrier
\bibliography{iras}{}

@ARTICLE{Sturm2023,
       author = {{Sturm}, J.~A. and {McClure}, M.~K. and {Bergner}, J.~B. and {Harsono}, D. and {Dartois}, E. and {Drozdovskaya}, M.~N. and {Ioppolo}, S. and {{\"O}berg}, K.~I. and {Law}, C.~J. and {Palumbo}, M.~E. and {Pendleton}, Y.~J. and {Rocha}, W.~R.~M. and {Terada}, H. and {Urso}, R.~G.},
        title = "{The edge-on protoplanetary disk HH 48 NE. II. Modeling ices and silicates}",
      journal = {\aap},
         year = 2023,
        month = sep,
       volume = {677},
          eid = {A18},
        pages = {A18},
          doi = {10.1051/0004-6361/202346053},
archivePrefix = {arXiv},
       eprint = {2305.02355},
 primaryClass = {astro-ph.EP},
       adsurl = {https://ui.adsabs.harvard.edu/abs/2023A&A...677A..18S}
}

@ARTICLE{Keane2001,
       author = {{Keane}, J.~V. and {Tielens}, A.~G.~G.~M. and {Boogert}, A.~C.~A. and {Schutte}, W.~A. and {Whittet}, D.~C.~B.},
        title = "{Ice absorption features in the 5-8 {\ensuremath{\mu}}m region toward embedded protostars}",
      journal = {\aap},
         year = 2001,
        month = sep,
       volume = {376},
        pages = {254-270},
          doi = {10.1051/0004-6361:20010936},
       adsurl = {https://ui.adsabs.harvard.edu/abs/2001A&A...376..254K}
}

@ARTICLE{Andrews2018,
       author = {{Andrews}, Sean M.},
        title = "{Observations of Protoplanetary Disk Structures}",
      journal = {\araa},
         year = 2020,
        month = aug,
       volume = {58},
        pages = {483-528},
          doi = {10.1146/annurev-astro-031220-010302},
archivePrefix = {arXiv},
       eprint = {2001.05007},
 primaryClass = {astro-ph.EP},
       adsurl = {https://ui.adsabs.harvard.edu/abs/2020ARA&A..58..483A}
}

@ARTICLE{Shu1987,
       author = {{Shu}, Frank H. and {Adams}, Fred C. and {Lizano}, Susana},
        title = "{Star formation in molecular clouds: observation and theory.}",
      journal = {\araa},
         year = 1987,
        month = jan,
       volume = {25},
        pages = {23-81},
          doi = {10.1146/annurev.aa.25.090187.000323},
       adsurl = {https://ui.adsabs.harvard.edu/abs/1987ARA&A..25...23S}
}

@ARTICLE{McKee2007,
       author = {{McKee}, Christopher F. and {Ostriker}, Eve C.},
        title = "{Theory of Star Formation}",
      journal = {\araa},
         year = 2007,
        month = sep,
       volume = {45},
       number = {1},
        pages = {565-687},
          doi = {10.1146/annurev.astro.45.051806.110602},
archivePrefix = {arXiv},
       eprint = {0707.3514},
 primaryClass = {astro-ph},
       adsurl = {https://ui.adsabs.harvard.edu/abs/2007ARA&A..45..565M}
}

@ARTICLE{Whitney2003,
       author = {{Whitney}, Barbara A. and {Wood}, Kenneth and {Bjorkman}, J.~E. and {Cohen}, Martin},
        title = "{Two-dimensional Radiative Transfer in Protostellar Envelopes. II. An Evolutionary Sequence}",
      journal = {\apj},
         year = 2003,
        month = dec,
       volume = {598},
       number = {2},
        pages = {1079-1099},
          doi = {10.1086/379068},
archivePrefix = {arXiv},
       eprint = {astro-ph/0309007},
 primaryClass = {astro-ph},
       adsurl = {https://ui.adsabs.harvard.edu/abs/2003ApJ...598.1079W}
}

@ARTICLE{Oberg2011,
       author = {{{\"O}berg}, Karin I. and {Murray-Clay}, Ruth and {Bergin}, Edwin A.},
        title = "{The Effects of Snowlines on C/O in Planetary Atmospheres}",
      journal = {\apjl},
         year = 2011,
        month = dec,
       volume = {743},
       number = {1},
          eid = {L16},
        pages = {L16},
          doi = {10.1088/2041-8205/743/1/L16},
archivePrefix = {arXiv},
       eprint = {1110.5567},
 primaryClass = {astro-ph.GA},
       adsurl = {https://ui.adsabs.harvard.edu/abs/2011ApJ...743L..16O}
}

@article{Henning2013,
  title={Chemistry in protoplanetary disks},
  author={Henning, Thomas and Semenov, Dmitry},
  journal={Chemical Reviews},
  volume={113},
  number={12},
  pages={9016--9042},
  year={2013},
  publisher={ACS Publications}
}

@ARTICLE{Pontoppidan2005,
       author = {{Pontoppidan}, Klaus M. and {Dullemond}, Cornelis P. and {van Dishoeck}, Ewine F. and {Blake}, Geoffrey A. and {Boogert}, Adwin C.~A. and {Evans}, II, Neal J. and {Kessler-Silacci}, Jacqueline E. and {Lahuis}, Fred},
        title = "{Ices in the Edge-on Disk CRBR 2422.8-3423: Spitzer Spectroscopy and Monte Carlo Radiative Transfer Modeling}",
      journal = {\apj},
         year = 2005,
        month = mar,
       volume = {622},
       number = {1},
        pages = {463-481},
          doi = {10.1086/427688},
archivePrefix = {arXiv},
       eprint = {astro-ph/0411367},
 primaryClass = {astro-ph},
       adsurl = {https://ui.adsabs.harvard.edu/abs/2005ApJ...622..463P}
}

@ARTICLE{Ossenkopf1994,
       author = {{Ossenkopf}, V. and {Henning}, Th.},
        title = "{Dust opacities for protostellar cores.}",
      journal = {\aap},
         year = 1994,
        month = nov,
       volume = {291},
        pages = {943-959},
       adsurl = {https://ui.adsabs.harvard.edu/abs/1994A&A...291..943O}
}

@ARTICLE{Whittet1983,
       author = {{Whittet}, D.~C.~B. and {Bode}, M.~F. and {Longmore}, A.~J. and {Baines}, D.~W.~T. and {Evans}, A.},
        title = "{Interstellar ice grains in the Taurus molecular clouds}",
      journal = {\nat},
         year = 1983,
        month = may,
       volume = {303},
       number = {5914},
        pages = {218-221},
          doi = {10.1038/303218a0},
       adsurl = {https://ui.adsabs.harvard.edu/abs/1983Natur.303..218W}
}

@ARTICLE{Gillett1973,
       author = {{Gillett}, F.~C. and {Forrest}, W.~J.},
        title = "{Spectra of the Becklin-Neugebauer point source and the Kleinmann-Low nebula from 2.8 to 13.5 microns.}",
      journal = {\apj},
         year = 1973,
        month = jan,
       volume = {179},
        pages = {483},
          doi = {10.1086/151888},
       adsurl = {https://ui.adsabs.harvard.edu/abs/1973ApJ...179..483G}
}

@ARTICLE{Dartois2024,
       author = {{Dartois}, E. and {Noble}, J.~A. and {Caselli}, P. and {Fraser}, H.~J. and {Jim{\'e}nez-Serra}, I. and {Mat{\'e}}, B. and {McClure}, M.~K. and {Melnick}, G.~J. and {Pendleton}, Y.~J. and {Shimonishi}, T. and {Smith}, Z.~L. and {Sturm}, J.~A. and {Taillard}, A. and {Wakelam}, V. and {Boogert}, A.~C.~A. and {Drozdovskaya}, M.~N. and {Erkal}, J. and {Harsono}, D. and {Herrero}, V.~J. and {Ioppolo}, S. and {Linnartz}, H. and {McGuire}, B.~A. and {Perotti}, G. and {Qasim}, D. and {Rocha}, W.~R.~M.},
        title = "{Spectroscopic sizing of interstellar icy grains with JWST}",
      journal = {Nature Astronomy},
         year = 2024,
        month = mar,
       volume = {8},
        pages = {359-367},
          doi = {10.1038/s41550-023-02155-x},
       adsurl = {https://ui.adsabs.harvard.edu/abs/2024NatAs...8..359D}
}

@ARTICLE{Evans1999,
       author = {{Evans}, II, Neal J.},
        title = "{Physical Conditions in Regions of Star Formation}",
      journal = {\araa},
         year = 1999,
        month = jan,
       volume = {37},
        pages = {311-362},
          doi = {10.1146/annurev.astro.37.1.311},
archivePrefix = {arXiv},
       eprint = {astro-ph/9905050},
 primaryClass = {astro-ph},
       adsurl = {https://ui.adsabs.harvard.edu/abs/1999ARA&A..37..311E}
}

@article{Weingartner2001,
doi = {10.1086/318651},
url = {https://dx.doi.org/10.1086/318651},
year = {2001},
month = {feb},
publisher = {},
volume = {548},
number = {1},
pages = {296},
author = {Weingartner, Joseph C. and Draine, B. T.},
title = {Dust Grain-Size Distributions and Extinction in the Milky Way,
Large Magellanic Cloud, and Small Magellanic Cloud},
journal = {The Astrophysical Journal}
}

@ARTICLE{Draine2003,
       author = {{Draine}, B.T.},
        title = "{Interstellar Dust Grains}",
      journal = {\araa},
         year = 2003,
        month = jan,
       volume = {41},
        pages = {241-289},
          doi = {10.1146/annurev.astro.41.011802.094840},
archivePrefix = {arXiv},
       eprint = {astro-ph/0304489},
 primaryClass = {astro-ph},
       adsurl = {https://ui.adsabs.harvard.edu/abs/2003ARA&A..41..241D}
}

@ARTICLE{Gibb2004,
       author = {{Gibb}, E.L. and {Whittet}, D.C.B. and {Boogert}, A.C.A. and {Tielens}, A.G.G.M.},
        title = "{Interstellar Ice: The Infrared Space Observatory Legacy}",
      journal = {\apjs},
         year = 2004,
        month = mar,
       volume = {151},
       number = {1},
        pages = {35-73},
          doi = {10.1086/381182},
       adsurl = {https://ui.adsabs.harvard.edu/abs/2004ApJS..151...35G}
}

@ARTICLE{Yang2022,
       author = {{Yang}, Yao-Lun and {Green}, Joel D. and {Pontoppidan}, Klaus M. and {Bergner}, Jennifer B. and {Cleeves}, L. Ilsedore and {Evans}, II, Neal J. and {Garrod}, Robin T. and {Jin}, Miwha and {Kim}, Chul Hwan and {Kim}, Jaeyeong and {Lee}, Jeong-Eun and {Sakai}, Nami and {Shingledecker}, Christopher N. and {Shope}, Brielle and {Tobin}, John J. and {van Dishoeck}, Ewine F.},
        title = "{CORINOS. I. JWST/MIRI Spectroscopy and Imaging of a Class 0 Protostar IRAS 15398{\textendash}3359}",
      journal = {\apjl},
         year = 2022,
        month = dec,
       volume = {941},
       number = {1},
          eid = {L13},
        pages = {L13},
          doi = {10.3847/2041-8213/aca289},
archivePrefix = {arXiv},
       eprint = {2208.10673},
 primaryClass = {astro-ph.SR},
       adsurl = {https://ui.adsabs.harvard.edu/abs/2022ApJ...941L..13Y}
}

@ARTICLE{Ulrich1976,
       author = {{Ulrich}, R.~K.},
        title = "{An infall model for the T Tauri phenomenon.}",
      journal = {\apj},
         year = 1976,
        month = dec,
       volume = {210},
        pages = {377-391},
          doi = {10.1086/154840},
       adsurl = {https://ui.adsabs.harvard.edu/abs/1976ApJ...210..377U}
}

@ARTICLE{Brunken2025,
       author = {{Brunken}, N.~G.~C. and {Boogert}, A.~C.~A. and {van Dishoeck}, E.~F. and {Evans}, N.~J. and {Poteet}, C.~A. and {Slavicinska}, K. and {Tychoniec}, L. and {Nazari}, P. and {Looney}, L.~W. and {Tyagi}, H. and {Narang}, M. and {Klaassen}, P. and {Yang}, Y. and {Kavanagh}, P.~J. and {Megeath}, S.~T. and {Ressler}, M.~E.},
        title = "{JWST observations of segregated (12) CO\_2 and (13) CO\_2 ices in protostellar envelopes}",
      journal = {ACS Earth and Space Chemistry},
         year = 2025,
        month = jul,
       volume = {9},
       number = {8},
        pages = {1992-2003},
          doi = {10.1021/acsearthspacechem.5c00037},
archivePrefix = {arXiv},
       eprint = {2505.14769},
 primaryClass = {astro-ph.SR},
       adsurl = {https://ui.adsabs.harvard.edu/abs/2025ESC.....9.1992B}
}

@ARTICLE{McClure2023,
       author = {{McClure}, M.~K. and {Rocha}, W.~R.~M. and {Pontoppidan}, K.~M. and {Crouzet}, N. and {Chu}, L.~E.~U. and {Dartois}, E. and {Lamberts}, T. and {Noble}, J.~A. and {Pendleton}, Y.~J. and {Perotti}, G. and {Qasim}, D. and {Rachid}, M.~G. and {Smith}, Z.~L. and {Sun}, Fengwu and {Beck}, Tracy L. and {Boogert}, A.~C.~A. and {Brown}, W.~A. and {Caselli}, P. and {Charnley}, S.~B. and {Cuppen}, Herma M. and {Dickinson}, H. and {Drozdovskaya}, M.~N. and {Egami}, E. and {Erkal}, J. and {Fraser}, H. and {Garrod}, R.~T. and {Harsono}, D. and {Ioppolo}, S. and {Jim{\'e}nez-Serra}, I. and {Jin}, M. and {J{\o}rgensen}, J.~K. and {Kristensen}, L.~E. and {Lis}, D.~C. and {McCoustra}, M.~R.~S. and {McGuire}, Brett A. and {Melnick}, G.~J. and {{\~A}-berg}, Karin I. and {Palumbo}, M.~E. and {Shimonishi}, T. and {Sturm}, J.~A. and {van Dishoeck}, E.~F. and {Linnartz}, H.},
        title = "{An Ice Age JWST inventory of dense molecular cloud ices}",
      journal = {Nature Astronomy},
         year = 2023,
        month = apr,
       volume = {7},
        pages = {431-443},
          doi = {10.1038/s41550-022-01875-w},
archivePrefix = {arXiv},
       eprint = {2301.09140},
 primaryClass = {astro-ph.GA},
       adsurl = {https://ui.adsabs.harvard.edu/abs/2023NatAs...7..431M}
}

@ARTICLE{Rocha2024,
       author = {{Rocha}, W.~R.~M. and {van Dishoeck}, E.~F. and {Ressler}, M.~E. and {van Gelder}, M.~L. and {Slavicinska}, K. and {Brunken}, N.~G.~C. and {Linnartz}, H. and {Ray}, T.~P. and {Beuther}, H. and {Caratti o Garatti}, A. and {Geers}, V. and {Kavanagh}, P.~J. and {Klaassen}, P.~D. and {Justtanont}, K. and {Chen}, Y. and {Francis}, L. and {Gieser}, C. and {Perotti}, G. and {Tychoniec}, {\L}. and {Barsony}, M. and {Majumdar}, L. and {le Gouellec}, V.~J.~M. and {Chu}, L.~E.~U. and {Lew}, B.~W.~P. and {Henning}, Th. and {Wright}, G.},
        title = "{JWST Observations of Young protoStars (JOYS+): Detecting icy complex organic molecules and ions. I. CH$_{4}$, SO$_{2}$, HCOO$^{{\ensuremath{-}}}$, OCN$^{{\ensuremath{-}}}$, H$_{2}$CO, HCOOH, CH$_{3}$CH$_{2}$OH, CH$_{3}$CHO, CH$_{3}$OCHO, and CH$_{3}$COOH}",
      journal = {\aap},
         year = 2024,
        month = mar,
       volume = {683},
          eid = {A124},
        pages = {A124},
          doi = {10.1051/0004-6361/202348427},
archivePrefix = {arXiv},
       eprint = {2312.06834},
 primaryClass = {astro-ph.SR},
       adsurl = {https://ui.adsabs.harvard.edu/abs/2024A&A...683A.124R}
}

@ARTICLE{Rocha2025,
       author = {{Rocha}, W.~R.~M. and {McClure}, M.~K. and {Sturm}, J.~A. and {Beck}, T.~L. and {Smith}, Z.~L. and {Dickinson}, H. and {Sun}, F. and {Egami}, E. and {Boogert}, A.~C.~A. and {Fraser}, H.~J. and {Dartois}, E. and {Jimenez-Serra}, I. and {Noble}, J.~A. and {Bergner}, J. and {Caselli}, P. and {Charnley}, S.~B. and {Chiar}, J. and {Chu}, L. and {Cooke}, I. and {Crouzet}, N. and {van Dishoeck}, E.~F. and {Drozdovskaya}, M.~N. and {Garrod}, R. and {Harsono}, D. and {Ioppolo}, S. and {Jin}, M. and {J{\o}rgensen}, J.~K. and {Lamberts}, T. and {Lis}, D.~C. and {Melnick}, G.~J. and {McGuire}, B.~A. and {{\"O}berg}, K.~I. and {Palumbo}, M.~E. and {Pendleton}, Y.~J. and {Perotti}, G. and {Qasim}, D. and {Shope}, B. and {Urso}, R.~G. and {Viti}, S. and {Linnartz}, H.},
        title = "{Ice inventory towards the protostar Ced 110 IRS4 observed with the James Webb Space Telescope: Results from the Early Release Science Ice Age program}",
      journal = {\aap},
         year = 2025,
        month = jan,
       volume = {693},
          eid = {A288},
        pages = {A288},
          doi = {10.1051/0004-6361/202451505},
archivePrefix = {arXiv},
       eprint = {2411.19651},
 primaryClass = {astro-ph.SR},
       adsurl = {https://ui.adsabs.harvard.edu/abs/2025A&A...693A.288R}
}

@ARTICLE{Boogert2015,
       author = {{Boogert}, A.~C. Adwin and {Gerakines}, Perry A. and {Whittet}, Douglas C.~B.},
        title = "{Observations of the icy universe.}",
      journal = {\araa},
         year = 2015,
        month = aug,
       volume = {53},
        pages = {541-581},
          doi = {10.1146/annurev-astro-082214-122348},
archivePrefix = {arXiv},
       eprint = {1501.05317},
 primaryClass = {astro-ph.GA},
       adsurl = {https://ui.adsabs.harvard.edu/abs/2015ARA&A..53..541B}
}

@ARTICLE{Boogert2008,
       author = {{Boogert}, A.~C.~A. and {Pontoppidan}, K.~M. and {Knez}, C. and {Lahuis}, F. and {Kessler-Silacci}, J. and {van Dishoeck}, E.~F. and {Blake}, G.~A. and {Augereau}, J. -C. and {Bisschop}, S.~E. and {Bottinelli}, S. and {Brooke}, T.~Y. and {Brown}, J. and {Crapsi}, A. and {Evans}, II, N.~J. and {Fraser}, H.~J. and {Geers}, V. and {Huard}, T.~L. and {J{\o}rgensen}, J.~K. and {{\"O}berg}, K.~I. and {Allen}, L.~E. and {Harvey}, P.~M. and {Koerner}, D.~W. and {Mundy}, L.~G. and {Padgett}, D.~L. and {Sargent}, A.~I. and {Stapelfeldt}, K.~R.},
        title = "{The c2d Spitzer Spectroscopic Survey of Ices around Low-Mass Young Stellar Objects. I. H$_{2}$O and the 5-8 {\ensuremath{\mu}}m Bands}",
      journal = {\apj},
         year = 2008,
        month = may,
       volume = {678},
       number = {2},
        pages = {985-1004},
          doi = {10.1086/533425},
archivePrefix = {arXiv},
       eprint = {0801.1167},
 primaryClass = {astro-ph},
       adsurl = {https://ui.adsabs.harvard.edu/abs/2008ApJ...678..985B}
}

@ARTICLE{Sturm2024,
       author = {{Sturm}, J.~A. and {McClure}, M.~K. and {Harsono}, D. and {Bergner}, J.~B. and {Dartois}, E. and {Boogert}, A.~C.~A. and {Cordiner}, M.~A. and {Drozdovskaya}, M.~N. and {Ioppolo}, S. and {Law}, C.~J. and {Lis}, D.~C. and {McGuire}, B.~A. and {Melnick}, G.~J. and {Noble}, J.~A. and {{\"O}berg}, K.~I. and {Palumbo}, M.~E. and {Pendleton}, Y.~J. and {Perotti}, G. and {Rocha}, W.~R.~M. and {Urso}, R.~G. and {van Dishoeck}, E.~F.},
        title = "{A JWST/MIRI analysis of the ice distribution and polycyclic aromatic hydrocarbon emission in the protoplanetary disk HH 48 NE}",
      journal = {\aap},
         year = 2024,
        month = sep,
       volume = {689},
          eid = {A92},
        pages = {A92},
          doi = {10.1051/0004-6361/202450865},
archivePrefix = {arXiv},
       eprint = {2407.09627},
 primaryClass = {astro-ph.EP},
       adsurl = {https://ui.adsabs.harvard.edu/abs/2024A&A...689A..92S}
}

@misc{Dullemond2012,
       author = {{Dullemond}, C.~P. and {Juhasz}, A. and {Pohl}, A. and {Sereshti}, F. and {Shetty}, R. and {Peters}, T. and {Commercon}, B. and {Flock}, M.},
        title = "{RADMC-3D: A multi-purpose radiative transfer tool}",
 howpublished = {Astrophysics Source Code Library, record ascl:1202.015},
         year = 2012,
        month = feb,
          eid = {ascl:1202.015},
       adsurl = {https://ui.adsabs.harvard.edu/abs/2012ascl.soft02015D}
}

@ARTICLE{Yang2017,
       author = {{Yang}, Yao-Lun and {Evans}, II, Neal J. and {Green}, Joel D. and {Dunham}, Michael M. and {J{\o}rgensen}, Jes K.},
        title = "{The Class 0 Protostar BHR71: Herschel Observations and Dust Continuum Models}",
      journal = {\apj},
         year = 2017,
        month = feb,
       volume = {835},
       number = {2},
          eid = {259},
        pages = {259},
          doi = {10.3847/1538-4357/835/2/259},
archivePrefix = {arXiv},
       eprint = {1701.00803},
 primaryClass = {astro-ph.SR},
       adsurl = {https://ui.adsabs.harvard.edu/abs/2017ApJ...835..259Y}
}

@ARTICLE{Hudgins1993,
       author = {{Hudgins}, D.~M. and {Sandford}, S.~A. and {Allamandola}, L.~J. and {Tielens}, A.~G.~G.~M.},
        title = "{Mid- and Far-Infrared Spectroscopy of Ices: Optical Constants and Integrated Absorbances}",
      journal = {\apjs},
         year = 1993,
        month = jun,
       volume = {86},
        pages = {713},
          doi = {10.1086/191796},
       adsurl = {https://ui.adsabs.harvard.edu/abs/1993ApJS...86..713H}
}

@ARTICLE{Warren2008,
       author = {{Warren}, Stephen G. and {Brandt}, Richard E.},
        title = "{Optical constants of ice from the ultraviolet to the microwave: A revised compilation}",
      journal = {Journal of Geophysical Research (Atmospheres)},
         year = 2008,
        month = jul,
       volume = {113},
       number = {D14},
          eid = {D14220},
        pages = {D14220},
          doi = {10.1029/2007JD009744},
       adsurl = {https://ui.adsabs.harvard.edu/abs/2008JGRD..11314220W}
}

@ARTICLE{Warren1986,
       author = {{Warren}, S.~G.},
        title = "{Optical constants of carbon dioxide ice}",
      journal = {\ao},
         year = 1986,
        month = aug,
       volume = {25},
        pages = {2650-2674},
          doi = {10.1364/AO.25.002650},
       adsurl = {https://ui.adsabs.harvard.edu/abs/1986ApOpt..25.2650W}
}

@ARTICLE{Palumbo2006,
       author = {{Palumbo}, M.~E. and {Baratta}, G.~A. and {Collings}, M.~P. and {McCoustra}, M.~R.~S.},
        title = "{The profile of the 2140 cm-1 solid CO band on different substrates}",
      journal = {Physical Chemistry Chemical Physics (Incorporating Faraday Transactions)},
         year = 2006,
        month = jan,
       volume = {8},
        pages = {279-284},
          doi = {10.1039/B509279E},
       adsurl = {https://ui.adsabs.harvard.edu/abs/2006PCCP....8..279P}
}

@ARTICLE{Pontoppidan2024,
       author = {{Pontoppidan}, Klaus M. and {Evans}, Neal and {Bergner}, Jennifer and {Yang}, Yao-Lun},
        title = "{A Constrained Dust Opacity for Models of Dense Clouds and Protostellar Envelopes}",
      journal = {Research Notes of the American Astronomical Society},
         year = 2024,
        month = mar,
       volume = {8},
       number = {3},
          eid = {68},
        pages = {68},
          doi = {10.3847/2515-5172/ad303f},
       adsurl = {https://ui.adsabs.harvard.edu/abs/2024RNAAS...8...68P}
}

@ARTICLE{Dominik2021,
       author = {{Dominik}, Carsten and {Min}, Michiel and {Tazaki}, Ryo},
        title = "{OpTool: Command-line driven tool for creating complex dust opacities}",
         year = 2021,
        month = apr,
      journal = {Astrophysics Source Code Library},
          eid = {ascl:2104.010},
        pages = {ascl:2104.010},
archivePrefix = {ascl},
       eprint = {2104.010},
      version = {1.9.4},
          url = {https://ascl.net/2104.010},
       adsurl = {https://ui.adsabs.harvard.edu/abs/2021ascl.soft04010D}
}

@ARTICLE{Heyer1989,
       author = {{Heyer}, Mark H. and {Graham}, J.~A.},
        title = "{Newborn Stars and Stellar Objects in Barnard 228}",
      journal = {\pasp},
         year = 1989,
        month = sep,
       volume = {101},
        pages = {816},
          doi = {10.1086/132502},
       adsurl = {https://ui.adsabs.harvard.edu/abs/1989PASP..101..816H}
}

@ARTICLE{Thieme2023,
       author = {{Thieme}, Travis J. and {Lai}, Shih-Ping and {Ohashi}, Nagayoshi and {Tobin}, John J. and {J{\o}rgensen}, Jes K. and {Sai}, Jinshi (Insa Choi) and {Aso}, Yusuke and {Williams}, Jonathan P. and {Yamato}, Yoshihide and {Aikawa}, Yuri and {de Gregorio-Monsalvo}, Itziar and {Han}, Ilseung and {Kwon}, Woojin and {Lee}, Chang Won and {Lee}, Jeong-Eun and {Li}, Zhi-Yun and {Lin}, Zhe-Yu Daniel and {Looney}, Leslie W. and {Narayanan}, Suchitra and {Phuong}, Nguyen Thi and {Plunkett}, Adele L. and {Santamar{\'\i}a-Miranda}, Alejandro and {Sharma}, Rajeeb and {Takakuwa}, Shigehisa and {Yen}, Hsi-Wei},
        title = "{Early Planet Formation in Embedded Disks (eDisk). VIII. A Small Protostellar Disk around the Extremely Low Mass and Young Class 0 Protostar IRAS 15398-3359}",
      journal = {\apj},
         year = 2023,
        month = nov,
       volume = {958},
       number = {1},
          eid = {60},
        pages = {60},
          doi = {10.3847/1538-4357/ad003a},
archivePrefix = {arXiv},
       eprint = {2310.12453},
 primaryClass = {astro-ph.EP},
       adsurl = {https://ui.adsabs.harvard.edu/abs/2023ApJ...958...60T}
}

@ARTICLE{Galli2020,
       author = {{Galli}, P.~A.~B. and {Bouy}, H. and {Olivares}, J. and {Miret-Roig}, N. and {Vieira}, R.~G. and {Sarro}, L.~M. and {Barrado}, D. and {Berihuete}, A. and {Bertout}, C. and {Bertin}, E. and {Cuillandre}, J. -C.},
        title = "{Lupus DANCe. Census of stars and 6D structure with Gaia-DR2 data}",
      journal = {\aap},
         year = 2020,
        month = nov,
       volume = {643},
          eid = {A148},
        pages = {A148},
          doi = {10.1051/0004-6361/202038717},
archivePrefix = {arXiv},
       eprint = {2010.00233},
 primaryClass = {astro-ph.SR},
       adsurl = {https://ui.adsabs.harvard.edu/abs/2020A&A...643A.148G}
}

@ARTICLE{Yen2017,
       author = {{Yen}, Hsi-Wei and {Koch}, Patrick M. and {Takakuwa}, Shigehisa and {Krasnopolsky}, Ruben and {Ohashi}, Nagayoshi and {Aso}, Yusuke},
        title = "{Signs of Early-stage Disk Growth Revealed with ALMA}",
      journal = {\apj},
         year = 2017,
        month = jan,
       volume = {834},
       number = {2},
          eid = {178},
        pages = {178},
          doi = {10.3847/1538-4357/834/2/178},
archivePrefix = {arXiv},
       eprint = {1611.08416},
 primaryClass = {astro-ph.SR},
       adsurl = {https://ui.adsabs.harvard.edu/abs/2017ApJ...834..178Y}
}

@ARTICLE{Okoda2018,
       author = {{Okoda}, Yuki and {Oya}, Yoko and {Sakai}, Nami and {Watanabe}, Yoshimasa and {J{\o}rgensen}, Jes K. and {Van Dishoeck}, Ewine F. and {Yamamoto}, Satoshi},
        title = "{The Co-evolution of Disks and Stars in Embedded Stages: The Case of the Very-low-mass Protostar IRAS 15398-3359}",
      journal = {\apjl},
         year = 2018,
        month = sep,
       volume = {864},
       number = {2},
          eid = {L25},
        pages = {L25},
          doi = {10.3847/2041-8213/aad8ba},
archivePrefix = {arXiv},
       eprint = {1809.00500},
 primaryClass = {astro-ph.SR},
       adsurl = {https://ui.adsabs.harvard.edu/abs/2018ApJ...864L..25O}
}

@ARTICLE{Jorgensen2013,
       author = {{J{\o}rgensen}, Jes K. and {Visser}, Ruud and {Sakai}, Nami and {Bergin}, Edwin A. and {Brinch}, Christian and {Harsono}, Daniel and {Lindberg}, Johan E. and {van Dishoeck}, Ewine F. and {Yamamoto}, Satoshi and {Bisschop}, Suzanne E. and {Persson}, Magnus V.},
        title = "{A Recent Accretion Burst in the Low-mass Protostar IRAS 15398-3359: ALMA Imaging of Its Related Chemistry}",
      journal = {\apjl},
         year = 2013,
        month = dec,
       volume = {779},
       number = {2},
          eid = {L22},
        pages = {L22},
          doi = {10.1088/2041-8205/779/2/L22},
archivePrefix = {arXiv},
       eprint = {1312.0724},
 primaryClass = {astro-ph.SR},
       adsurl = {https://ui.adsabs.harvard.edu/abs/2013ApJ...779L..22J}
}

@ARTICLE{Okoda2023,
       author = {{Okoda}, Yuki and {Oya}, Yoko and {Francis}, Logan and {Johnstone}, Doug and {Ceccarelli}, Cecilia and {Codella}, Claudio and {Chandler}, Claire J. and {Sakai}, Nami and {Aikawa}, Yuri and {Alves}, Felipe O. and {Herbst}, Eric and {Maureira}, Mar{\'\i}a Jos{\'e} and {Bouvier}, Mathilde and {Caselli}, Paola and {Choudhury}, Spandan and {De Simone}, Marta and {J{\'\i}menez-Serra}, Izaskun and {Pineda}, Jaime and {Yamamoto}, Satoshi},
        title = "{FAUST. VII. Detection of a Hot Corino in the Prototypical Warm Carbon-chain Chemistry Source IRAS 15398-3359}",
      journal = {\apj},
         year = 2023,
        month = may,
       volume = {948},
       number = {2},
          eid = {127},
        pages = {127},
          doi = {10.3847/1538-4357/acc1e5},
archivePrefix = {arXiv},
       eprint = {2303.03564},
 primaryClass = {astro-ph.SR},
       adsurl = {https://ui.adsabs.harvard.edu/abs/2023ApJ...948..127O}
}

@ARTICLE{Bottinelli2010,
       author = {{Bottinelli}, Sandrine and {Boogert}, A.~C. Adwin and {Bouwman}, Jordy and {Beckwith}, Martha and {van Dishoeck}, Ewine F. and {{\"O}berg}, Karin I. and {Pontoppidan}, Klaus M. and {Linnartz}, Harold and {Blake}, Geoffrey A. and {Evans}, II, Neal J. and {Lahuis}, Fred},
        title = "{The c2d Spitzer Spectroscopic Survey of Ices Around Low-mass Young Stellar Objects. IV. NH$_{3}$ and CH$_{3}$OH}",
      journal = {\apj},
         year = 2010,
        month = aug,
       volume = {718},
       number = {2},
        pages = {1100-1117},
          doi = {10.1088/0004-637X/718/2/1100},
archivePrefix = {arXiv},
       eprint = {1005.2225},
 primaryClass = {astro-ph.GA},
       adsurl = {https://ui.adsabs.harvard.edu/abs/2010ApJ...718.1100B}
}

@ARTICLE{Pontoppidan2008,
       author = {{Pontoppidan}, Klaus M. and {Boogert}, A.~C.~A. and {Fraser}, Helen J. and {van Dishoeck}, Ewine F. and {Blake}, Geoffrey A. and {Lahuis}, Fred and {{\"O}berg}, Karin I. and {Evans}, II, Neal J. and {Salyk}, Colette},
        title = "{The c2d Spitzer Spectroscopic Survey of Ices around Low-Mass Young Stellar Objects. II. CO$_{2}$}",
      journal = {\apj},
         year = 2008,
        month = may,
       volume = {678},
       number = {2},
        pages = {1005-1031},
          doi = {10.1086/533431},
archivePrefix = {arXiv},
       eprint = {0711.4616},
 primaryClass = {astro-ph},
       adsurl = {https://ui.adsabs.harvard.edu/abs/2008ApJ...678.1005P}
}

@ARTICLE{Oberg2008,
       author = {{{\"O}berg}, Karin I. and {Boogert}, A.~C. Adwin and {Pontoppidan}, Klaus M. and {Blake}, Geoffrey A. and {Evans}, Neal J. and {Lahuis}, Fred and {van Dishoeck}, Ewine F.},
        title = "{The c2d Spitzer Spectroscopic Survey of Ices around Low-Mass Young Stellar Objects. III. CH$_{4}$}",
      journal = {\apj},
         year = 2008,
        month = may,
       volume = {678},
       number = {2},
        pages = {1032-1041},
          doi = {10.1086/533432},
archivePrefix = {arXiv},
       eprint = {0801.1223},
 primaryClass = {astro-ph},
       adsurl = {https://ui.adsabs.harvard.edu/abs/2008ApJ...678.1032O}
}

@ARTICLE{Salyk2024,
       author = {{Salyk}, Colette and {Yang}, Yao-Lun and {Pontoppidan}, Klaus M. and {Bergner}, Jennifer B. and {Okoda}, Yuki and {Kim}, Jaeyeong and {Evans}, Neal J. and {Cleeves}, Ilsedore and {van Dishoeck}, Ewine F. and {Garrod}, Robin T. and {Green}, Joel D.},
        title = "{CORINOS. II. JWST-MIRI Detection of Warm Molecular Gas from an Embedded, Disk-bearing Protostar}",
      journal = {\apj},
         year = 2024,
        month = oct,
       volume = {974},
       number = {1},
          eid = {97},
        pages = {97},
          doi = {10.3847/1538-4357/ad62fe},
archivePrefix = {arXiv},
       eprint = {2407.15303},
 primaryClass = {astro-ph.SR},
       adsurl = {https://ui.adsabs.harvard.edu/abs/2024ApJ...974...97S}
}

@ARTICLE{Meyer1998,
       author = {{Meyer}, David M. and {Jura}, M. and {Cardelli}, Jason A.},
        title = "{The Definitive Abundance of Interstellar Oxygen}",
      journal = {\apj},
         year = 1998,
        month = jan,
       volume = {493},
       number = {1},
        pages = {222-229},
          doi = {10.1086/305128},
archivePrefix = {arXiv},
       eprint = {astro-ph/9710163},
 primaryClass = {astro-ph},
       adsurl = {https://ui.adsabs.harvard.edu/abs/1998ApJ...493..222M}
}

@ARTICLE{Chen2024,
       author = {{Chen}, Y. and {Rocha}, W.~R.~M. and {van Dishoeck}, E.~F. and {van Gelder}, M.~L. and {Nazari}, P. and {Slavicinska}, K. and {Francis}, L. and {Tabone}, B. and {Ressler}, M.~E. and {Klaassen}, P.~D. and {Beuther}, H. and {Boogert}, A.~C.~A. and {Gieser}, C. and {Kavanagh}, P.~J. and {Perotti}, G. and {Le Gouellec}, V.~J.~M. and {Majumdar}, L. and {G{\"u}del}, M. and {Henning}, Th.},
        title = "{JOYS+: The link between the ice and gas of complex organic molecules: Comparing JWST and ALMA data of two low-mass protostars}",
      journal = {\aap},
         year = 2024,
        month = oct,
       volume = {690},
          eid = {A205},
        pages = {A205},
          doi = {10.1051/0004-6361/202450706},
archivePrefix = {arXiv},
       eprint = {2407.20066},
 primaryClass = {astro-ph.GA},
       adsurl = {https://ui.adsabs.harvard.edu/abs/2024A&A...690A.205C}
}

@ARTICLE{Boogert2000,
       author = {{Boogert}, A.~C.~A. and {Ehrenfreund}, P. and {Gerakines}, P.~A. and {Tielens}, A.~G.~G.~M. and {Whittet}, D.~C.~B. and {Schutte}, W.~A. and {van Dishoeck}, E.~F. and {de Graauw}, Th. and {Decin}, L. and {Prusti}, T.},
        title = "{ISO-SWS observations of interstellar solid $^{13}$CO$_{2}$: heated ice and the Galactic $^{12}$C/$^{13}$C abundance ratio}",
      journal = {\aap},
         year = 2000,
        month = jan,
       volume = {353},
        pages = {349-362},
          doi = {10.48550/arXiv.astro-ph/9909477},
archivePrefix = {arXiv},
       eprint = {astro-ph/9909477},
 primaryClass = {astro-ph},
       adsurl = {https://ui.adsabs.harvard.edu/abs/2000A&A...353..349B}
}

@ARTICLE{Slavicinska2024,
       author = {{Slavicinska}, Katerina and {van Dishoeck}, Ewine F. and {Tychoniec}, {\L}ukasz and {Nazari}, Pooneh and {Rubinstein}, Adam E. and {Gutermuth}, Robert and {Tyagi}, Himanshu and {Chen}, Yuan and {Brunken}, Nashanty G.~C. and {Rocha}, Will R.~M. and {Manoj}, P. and {Narang}, Mayank and {Megeath}, S. Thomas and {Yang}, Yao-Lun and {Looney}, Leslie W. and {Tobin}, John J. and {Beuther}, Henrik and {Bourke}, Tyler L. and {Linnartz}, Harold and {Federman}, Samuel and {Watson}, Dan M. and {Linz}, Hendrik},
        title = "{JWST detections of amorphous and crystalline HDO ice toward massive protostars}",
      journal = {\aap},
         year = 2024,
        month = aug,
       volume = {688},
          eid = {A29},
        pages = {A29},
          doi = {10.1051/0004-6361/202449785},
archivePrefix = {arXiv},
       eprint = {2404.15399},
 primaryClass = {astro-ph.SR},
       adsurl = {https://ui.adsabs.harvard.edu/abs/2024A&A...688A..29S}
}

@ARTICLE{Slavicinska2025,
       author = {{Slavicinska}, Katerina and {Tychoniec}, {\L}ukasz and {Navarro}, Mar{\'\i}a Gabriela and {van Dishoeck}, Ewine F. and {Tobin}, John J. and {van Gelder}, Martijn L. and {Chen}, Yuan and {Boogert}, A.~C. Adwin and {Drechsler}, W. Blake and {Beuther}, Henrik and {Caratti o Garatti}, Alessio and {Megeath}, S. Thomas and {Klaassen}, Pamela and {Looney}, Leslie W. and {Kavanagh}, Patrick J. and {Brunken}, Nashanty G.~C. and {Sheehan}, Patrick and {Fischer}, William J.},
        title = "{HDO Ice Detected toward an Isolated Low-mass Protostar with JWST}",
      journal = {\apjl},
         year = 2025,
        month = jun,
       volume = {986},
       number = {2},
          eid = {L19},
        pages = {L19},
          doi = {10.3847/2041-8213/addb45},
archivePrefix = {arXiv},
       eprint = {2505.14686},
 primaryClass = {astro-ph.SR},
       adsurl = {https://ui.adsabs.harvard.edu/abs/2025ApJ...986L..19S}
}

@Article{Hunter2007,
  Author    = {Hunter, J. D.},
  Title     = {Matplotlib: A 2D graphics environment},
  Journal   = {Computing in Science \& Engineering},
  Volume    = {9},
  Number    = {3},
  Pages     = {90--95},
  publisher = {IEEE COMPUTER SOC},
  doi       = {10.1109/MCSE.2007.55},
  year      = 2007
}

@Article{harris2020,
 title         = {Array programming with {NumPy}},
 author        = {Charles R. Harris and K. Jarrod Millman and St{\'{e}}fan J.
                 van der Walt and Ralf Gommers and Pauli Virtanen and David
                 Cournapeau and Eric Wieser and Julian Taylor and Sebastian
                 Berg and Nathaniel J. Smith and Robert Kern and Matti Picus
                 and Stephan Hoyer and Marten H. van Kerkwijk and Matthew
                 Brett and Allan Haldane and Jaime Fern{\'{a}}ndez del
                 R{\'{i}}o and Mark Wiebe and Pearu Peterson and Pierre
                 G{\'{e}}rard-Marchant and Kevin Sheppard and Tyler Reddy and
                 Warren Weckesser and Hameer Abbasi and Christoph Gohlke and
                 Travis E. Oliphant},
 year          = {2020},
 journal       = {Nature},
 volume        = {585},
 number        = {7825},
 pages         = {357--362},
 doi           = {10.1038/s41586-020-2649-2},
 publisher     = {Springer Science and Business Media {LLC}},
 url           = {https://doi.org/10.1038/s41586-020-2649-2}
}

@misc{pandas2020,
    author       = {The pandas development team},
    title        = {pandas-dev/pandas: Pandas},
    month        = feb,
    year         = 2020,
    publisher    = {Zenodo},
    version      = {latest},
    doi          = {10.5281/zenodo.3509134},
    url          = {https://doi.org/10.5281/zenodo.3509134}
}

@ARTICLE{Green2016,
       author = {{Green}, Joel D. and {Yang}, Yao-Lun and {Evans}, II, Neal J. and {Karska}, Agata and {Herczeg}, Gregory and {van Dishoeck}, Ewine F. and {Lee}, Jeong-Eun and {Larson}, Rebecca L. and {Bouwman}, Jeroen},
        title = "{The CDF Archive: Herschel PACS and SPIRE Spectroscopic Data Pipeline and Products for Protostars and Young Stellar Objects}",
      journal = {\aj},
         year = 2016,
        month = mar,
       volume = {151},
       number = {3},
          eid = {75},
        pages = {75},
          doi = {10.3847/0004-6256/151/3/75},
archivePrefix = {arXiv},
       eprint = {1601.05028},
 primaryClass = {astro-ph.IM},
       adsurl = {https://ui.adsabs.harvard.edu/abs/2016AJ....151...75G}
}

@ARTICLE{Terebey1984,
       author = {{Terebey}, S. and {Shu}, F.~H. and {Cassen}, P.},
        title = "{The collapse of the cores of slowly rotating isothermal clouds}",
      journal = {\apj},
         year = 1984,
        month = nov,
       volume = {286},
        pages = {529-551},
          doi = {10.1086/162628},
       adsurl = {https://ui.adsabs.harvard.edu/abs/1984ApJ...286..529T}
}

@ARTICLE{Li2001,
       author = {{Li}, Aigen and {Draine}, B.~T.},
        title = "{Infrared Emission from Interstellar Dust. II. The Diffuse Interstellar Medium}",
      journal = {\apj},
         year = 2001,
        month = jun,
       volume = {554},
       number = {2},
        pages = {778-802},
          doi = {10.1086/323147},
archivePrefix = {arXiv},
       eprint = {astro-ph/0011319},
 primaryClass = {astro-ph},
       adsurl = {https://ui.adsabs.harvard.edu/abs/2001ApJ...554..778L}
}

@ARTICLE{Shirley2000,
       author = {{Shirley}, Yancy L. and {Evans}, II, Neal J. and {Rawlings}, Jonathan M.~C. and {Gregersen}, Erik M.},
        title = "{Tracing the Mass during Low-Mass Star Formation. I. Submillimeter Continuum Observations}",
      journal = {\apjs},
         year = 2000,
        month = nov,
       volume = {131},
       number = {1},
        pages = {249-271},
          doi = {10.1086/317358},
archivePrefix = {arXiv},
       eprint = {astro-ph/0006183},
 primaryClass = {astro-ph},
       adsurl = {https://ui.adsabs.harvard.edu/abs/2000ApJS..131..249S}
}

@ARTICLE{Draine2003astrosil,
       author = {{Draine}, B.~T.},
        title = "{Scattering by Interstellar Dust Grains. I. Optical and Ultraviolet}",
      journal = {\apj},
         year = 2003,
        month = dec,
       volume = {598},
       number = {2},
        pages = {1017-1025},
          doi = {10.1086/379118},
archivePrefix = {arXiv},
       eprint = {astro-ph/0304060},
 primaryClass = {astro-ph},
       adsurl = {https://ui.adsabs.harvard.edu/abs/2003ApJ...598.1017D}
}

@ARTICLE{Zubko1996,
       author = {{Zubko}, V.~G. and {Mennella}, V. and {Colangeli}, L. and {Bussoletti}, E.},
        title = "{Optical constants of cosmic carbon analogue grains - I. Simulation of clustering by a modified continuous distribution of ellipsoids}",
      journal = {\mnras},
         year = 1996,
        month = oct,
       volume = {282},
       number = {4},
        pages = {1321-1329},
          doi = {10.1093/mnras/282.4.1321},
       adsurl = {https://ui.adsabs.harvard.edu/abs/1996MNRAS.282.1321Z}
}

@ARTICLE{Draine2003graph,
       author = {{Draine}, B.~T.},
        title = "{Scattering by Interstellar Dust Grains. II. X-Rays}",
      journal = {\apj},
         year = 2003,
        month = dec,
       volume = {598},
       number = {2},
        pages = {1026-1037},
          doi = {10.1086/379123},
archivePrefix = {arXiv},
       eprint = {astro-ph/0308251},
 primaryClass = {astro-ph},
       adsurl = {https://ui.adsabs.harvard.edu/abs/2003ApJ...598.1026D}
}

@ARTICLE{Evans2009,
       author = {{Evans}, II, Neal J. and {Dunham}, Michael M. and {J{\o}rgensen}, Jes K. and {Enoch}, Melissa L. and {Mer{\'\i}n}, Bruno and {van Dishoeck}, Ewine F. and {Alcal{\'a}}, Juan M. and {Myers}, Philip C. and {Stapelfeldt}, Karl R. and {Huard}, Tracy L. and {Allen}, Lori E. and {Harvey}, Paul M. and {van Kempen}, Tim and {Blake}, Geoffrey A. and {Koerner}, David W. and {Mundy}, Lee G. and {Padgett}, Deborah L. and {Sargent}, Anneila I.},
        title = "{The Spitzer c2d Legacy Results: Star-Formation Rates and Efficiencies; Evolution and Lifetimes}",
      journal = {\apjs},
         year = 2009,
        month = apr,
       volume = {181},
       number = {2},
        pages = {321-350},
          doi = {10.1088/0067-0049/181/2/321},
archivePrefix = {arXiv},
       eprint = {0811.1059},
 primaryClass = {astro-ph},
       adsurl = {https://ui.adsabs.harvard.edu/abs/2009ApJS..181..321E}
}

@ARTICLE{Skrutskie2006,
       author = {{Skrutskie}, M.~F. and {Cutri}, R.~M. and {Stiening}, R. and {Weinberg}, M.~D. and {Schneider}, S. and {Carpenter}, J.~M. and {Beichman}, C. and {Capps}, R. and {Chester}, T. and {Elias}, J. and {Huchra}, J. and {Liebert}, J. and {Lonsdale}, C. and {Monet}, D.~G. and {Price}, S. and {Seitzer}, P. and {Jarrett}, T. and {Kirkpatrick}, J.~D. and {Gizis}, J.~E. and {Howard}, E. and {Evans}, T. and {Fowler}, J. and {Fullmer}, L. and {Hurt}, R. and {Light}, R. and {Kopan}, E.~L. and {Marsh}, K.~A. and {McCallon}, H.~L. and {Tam}, R. and {Van Dyk}, S. and {Wheelock}, S.},
        title = "{The Two Micron All Sky Survey (2MASS)}",
      journal = {\aj},
         year = 2006,
        month = feb,
       volume = {131},
       number = {2},
        pages = {1163-1183},
          doi = {10.1086/498708},
       adsurl = {https://ui.adsabs.harvard.edu/abs/2006AJ....131.1163S}
}

@ARTICLE{Andre2010,
       author = {{Andr{\'e}}, Ph. and {Men'shchikov}, A. and {Bontemps}, S. and {K{\"o}nyves}, V. and {Motte}, F. and {Schneider}, N. and {Didelon}, P. and {Minier}, V. and {Saraceno}, P. and {Ward-Thompson}, D. and {di Francesco}, J. and {White}, G. and {Molinari}, S. and {Testi}, L. and {Abergel}, A. and {Griffin}, M. and {Henning}, Th. and {Royer}, P. and {Mer{\'\i}n}, B. and {Vavrek}, R. and {Attard}, M. and {Arzoumanian}, D. and {Wilson}, C.~D. and {Ade}, P. and {Aussel}, H. and {Baluteau}, J.-P. and {Benedettini}, M. and {Bernard}, J.-Ph. and {Blommaert}, J.~A.~D.~L. and {Cambr{\'e}sy}, L. and {Cox}, P. and {di Giorgio}, A. and {Hargrave}, P. and {Hennemann}, M. and {Huang}, M. and {Kirk}, J. and {Krause}, O. and {Launhardt}, R. and {Leeks}, S. and {Le Pennec}, J. and {Li}, J.~Z. and {Martin}, P.~G. and {Maury}, A. and {Olofsson}, G. and {Omont}, A. and {Peretto}, N. and {Pezzuto}, S. and {Prusti}, T. and {Roussel}, H. and {Russeil}, D. and {Sauvage}, M. and {Sibthorpe}, B. and {Sicilia-Aguilar}, A. and {Spinoglio}, L. and {Waelkens}, C. and {Woodcraft}, A. and {Zavagno}, A.},
        title = "{From filamentary clouds to prestellar cores to the stellar IMF: Initial highlights from the Herschel Gould Belt Survey}",
      journal = {\aap},
         year = 2010,
        month = jul,
       volume = {518},
          eid = {L102},
        pages = {L102},
          doi = {10.1051/0004-6361/201014666},
archivePrefix = {arXiv},
       eprint = {1005.2618},
 primaryClass = {astro-ph.GA},
       adsurl = {https://ui.adsabs.harvard.edu/abs/2010A&A...518L.102A}
}

@article{nurnberger1997,
  title={A 1.3 mm dust continuum survey of Halpha selected T Tauri stars in Lupus.},
  author={N{\"u}rnberger, Dieter and Chini, Rolf and Zinnecker, Hans},
  journal={Astronomy and Astrophysics, v. 324, p. 1036-1045},
  volume={324},
  pages={1036--1045},
  year={1997}
}

@ARTICLE{Minissale2022,
       author = {{Minissale}, Marco and {Aikawa}, Yuri and {Bergin}, Edwin and {Bertin}, Mathieu and {Brown}, Wendy A. and {Cazaux}, Stephanie and {Charnley}, Steven B. and {Coutens}, Audrey and {Cuppen}, Herma M. and {Guzman}, Victoria and {Linnartz}, Harold and {McCoustra}, Martin R.~S. and {Rimola}, Albert and {Schrauwen}, Johanna G.~M. and {Toubin}, Celine and {Ugliengo}, Piero and {Watanabe}, Naoki and {Wakelam}, Valentine and {Dulieu}, Francois},
        title = "{Thermal Desorption of Interstellar Ices: A Review on the Controlling Parameters and Their Implications from Snowlines to Chemical Complexity}",
      journal = {ACS Earth and Space Chemistry},
         year = 2022,
        month = mar,
       volume = {6},
       number = {3},
        pages = {597-630},
          doi = {10.1021/acsearthspacechem.1c00357},
archivePrefix = {arXiv},
       eprint = {2201.07512},
 primaryClass = {astro-ph.GA},
       adsurl = {https://ui.adsabs.harvard.edu/abs/2022ESC.....6..597M}
}

@ARTICLE{Rayalacheruvu2025,
       author = {{Rayalacheruvu}, Prathap and {Majumdar}, Liton and {Rocha}, W.~R.~M. and {Ressler}, Michael E. and {Ranjan Giri}, Pabitra and {Maitrey}, S. and {Willacy}, K. and {Lis}, D.~C. and {Chen}, Y. and {Klaassen}, P.~D.},
        title = "{Expanding the Ice Inventory of NGC 1333 IRAS 2A with INDRA using JWST Observations: Tracing Organic Refractories and Beyond}",
      journal = {arXiv e-prints},
         year = 2025,
        month = jun,
          eid = {arXiv:2506.15358},
        pages = {arXiv:2506.15358},
          doi = {10.48550/arXiv.2506.15358},
archivePrefix = {arXiv},
       eprint = {2506.15358},
 primaryClass = {astro-ph.GA},
       adsurl = {https://ui.adsabs.harvard.edu/abs/2025arXiv250615358R}
}

@ARTICLE{Bouilloud2015,
       author = {{Bouilloud}, M. and {Fray}, N. and {B{\'e}nilan}, Y. and {Cottin}, H. and {Gazeau}, M.-C. and {Jolly}, A.},
        title = "{Bibliographic review and new measurements of the infrared band strengths of pure molecules at 25 K: H$_{2}$O, CO$_{2}$, CO, CH$_{4}$, NH$_{3}$, CH$_{3}$OH, HCOOH and H$_{2}$CO}",
      journal = {\mnras},
         year = 2015,
        month = aug,
       volume = {451},
       number = {2},
        pages = {2145-2160},
          doi = {10.1093/mnras/stv1021},
       adsurl = {https://ui.adsabs.harvard.edu/abs/2015MNRAS.451.2145B}
}

@ARTICLE{Dohnalek2003,
       author = {{Dohn{\'a}lek}, Z. and {Kimmel}, Greg A. and {Ayotte}, Patrick and {Smith}, R. Scott and {Kay}, Bruce D.},
        title = "{The deposition angle-dependent density of amorphous solid water films}",
      journal = {\jcp},
         year = 2003,
        month = jan,
       volume = {118},
       number = {1},
        pages = {364-372},
          doi = {10.1063/1.1525805},
       adsurl = {https://ui.adsabs.harvard.edu/abs/2003JChPh.118..364D}
}

@ARTICLE{Vazzano2021,
       author = {{Vazzano}, M.~M. and {Fern{\'a}ndez-L{\'o}pez}, M. and {Plunkett}, A. and {de Gregorio-Monsalvo}, I. and {Santamar{\'\i}a-Miranda}, A. and {Takahashi}, S. and {Lopez}, C.},
        title = "{Outflows, envelopes, and disks as evolutionary indicators in Lupus young stellar objects}",
      journal = {\aap},
         year = 2021,
        month = apr,
       volume = {648},
          eid = {A41},
        pages = {A41},
          doi = {10.1051/0004-6361/202039228},
archivePrefix = {arXiv},
       eprint = {2101.05330},
 primaryClass = {astro-ph.SR},
       adsurl = {https://ui.adsabs.harvard.edu/abs/2021A&A...648A..41V}
}

@ARTICLE{Okoda2025,
       author = {{Okoda}, Yuki and {Yang}, Yao-Lun and {Evans}, II, Neal J. and {Kim}, Jaeyeong and {Jin}, Mihwa and {Garrod}, Robin T. and {Francis}, Logan and {Johnstone}, Doug and {Ceccarelli}, Cecilia and {Codella}, Claudio and {Chandler}, Claire J. and {Yamamoto}, Satoshi and {Sakai}, Nami},
        title = "{CORINOS. III. Outflow Shocked Regions of the Low-mass Protostellar Source IRAS 15398{\textendash}3359 with JWST and ALMA}",
      journal = {\apj},
         year = 2025,
        month = apr,
       volume = {982},
       number = {2},
          eid = {149},
        pages = {149},
          doi = {10.3847/1538-4357/adb83f},
archivePrefix = {arXiv},
       eprint = {2503.03050},
 primaryClass = {astro-ph.SR},
       adsurl = {https://ui.adsabs.harvard.edu/abs/2025ApJ...982..149O}
}

@ARTICLE{Yang2020,
       author = {{Yang}, Yao-Lun and {Evans}, II, Neal J. and {Smith}, Aaron and {Lee}, Jeong-Eun and {Tobin}, John J. and {Terebey}, Susan and {Calcutt}, Hannah and {J{\o}rgensen}, Jes K. and {Green}, Joel D. and {Bourke}, Tyler L.},
        title = "{Constraining the Infalling Envelope Models of Embedded Protostars: BHR 71 and Its Hot Corino}",
      journal = {\apj},
         year = 2020,
        month = mar,
       volume = {891},
       number = {1},
          eid = {61},
        pages = {61},
          doi = {10.3847/1538-4357/ab7201},
archivePrefix = {arXiv},
       eprint = {2002.01478},
 primaryClass = {astro-ph.SR},
       adsurl = {https://ui.adsabs.harvard.edu/abs/2020ApJ...891...61Y}
}

@ARTICLE{Brunken2024,
       author = {{Brunken}, N.~G.~C. and {van Dishoeck}, E.~F. and {Slavicinska}, K. and {le Gouellec}, V.~J.~M. and {Rocha}, W.~R.~M. and {Francis}, L. and {Tychoniec}, L. and {van Gelder}, M.~L. and {Navarro}, M.~G. and {Boogert}, A.~C.~A. and {Kavanagh}, P.~J. and {Nazari}, P. and {Greene}, T. and {Ressler}, M.~E. and {Majumdar}, L.},
        title = "{JOYS+ study of solid-state $^{12}$C/$^{13}$C isotope ratios in protostellar envelopes: Observations of CO and CO$_{2}$ ice with the James Webb Space Telescope}",
      journal = {\aap},
         year = 2024,
        month = dec,
       volume = {692},
          eid = {A163},
        pages = {A163},
          doi = {10.1051/0004-6361/202451794},
archivePrefix = {arXiv},
       eprint = {2409.17237},
 primaryClass = {astro-ph.SR},
       adsurl = {https://ui.adsabs.harvard.edu/abs/2024A&A...692A.163B}
}

@ARTICLE{Arabhavi2022,
       author = {{Arabhavi}, Aditya M. and {Woitke}, Peter and {Cazaux}, St{\'e}phanie M. and {Kamp}, Inga and {Rab}, Christian and {Thi}, Wing-Fai},
        title = "{Ices in planet-forming disks: Self-consistent ice opacities in disk models}",
      journal = {\aap},
         year = 2022,
        month = oct,
       volume = {666},
          eid = {A139},
        pages = {A139},
          doi = {10.1051/0004-6361/202141825},
archivePrefix = {arXiv},
       eprint = {2208.12739},
 primaryClass = {astro-ph.EP},
       adsurl = {https://ui.adsabs.harvard.edu/abs/2022A&A...666A.139A}
}

@ARTICLE{Sakai2009,
       author = {{Sakai}, Nami and {Sakai}, Takeshi and {Hirota}, Tomoya and {Burton}, Michael and {Yamamoto}, Satoshi},
        title = "{Discovery of the Second Warm Carbon-Chain-Chemistry Source, IRAS15398 - 3359 in Lupus}",
      journal = {\apj},
         year = 2009,
        month = may,
       volume = {697},
       number = {1},
        pages = {769-786},
          doi = {10.1088/0004-637X/697/1/769},
       adsurl = {https://ui.adsabs.harvard.edu/abs/2009ApJ...697..769S}
}

@ARTICLE{Osorio2003,
       author = {{Osorio}, Mayra and {D'Alessio}, Paola and {Muzerolle}, James and {Calvet}, Nuria and {Hartmann}, Lee},
        title = "{A Comprehensive Study of the L1551 IRS 5 Binary System}",
      journal = {\apj},
         year = 2003,
        month = apr,
       volume = {586},
       number = {2},
        pages = {1148-1161},
          doi = {10.1086/367695},
archivePrefix = {arXiv},
       eprint = {astro-ph/0212074},
 primaryClass = {astro-ph},
       adsurl = {https://ui.adsabs.harvard.edu/abs/2003ApJ...586.1148O}
}

@ARTICLE{Shirley2002,
       author = {{Shirley}, Yancy L. and {Evans}, II, Neal J. and {Rawlings}, Jonathan M.~C.},
        title = "{Tracing the Mass during Low-Mass Star Formation. III. Models of the Submillimeter Dust Continuum Emission from Class 0 Protostars}",
      journal = {\apj},
         year = 2002,
        month = aug,
       volume = {575},
       number = {1},
        pages = {337-353},
          doi = {10.1086/341286},
archivePrefix = {arXiv},
       eprint = {astro-ph/0204024},
 primaryClass = {astro-ph},
       adsurl = {https://ui.adsabs.harvard.edu/abs/2002ApJ...575..337S}
}

@ARTICLE{Min2005,
       author = {{Min}, M. and {Hovenier}, J.~W. and {de Koter}, A.},
        title = "{Modeling optical properties of cosmic dust grains using a distribution of hollow spheres}",
      journal = {\aap},
         year = 2005,
        month = mar,
       volume = {432},
       number = {3},
        pages = {909-920},
          doi = {10.1051/0004-6361:20041920},
archivePrefix = {arXiv},
       eprint = {astro-ph/0503068},
 primaryClass = {astro-ph},
       adsurl = {https://ui.adsabs.harvard.edu/abs/2005A&A...432..909M}
}

@ARTICLE{Shakura1973,
       author = {{Shakura}, N.~I. and {Sunyaev}, R.~A.},
        title = "{Black holes in binary systems. Observational appearance.}",
      journal = {\aap},
         year = 1973,
        month = jan,
       volume = {24},
        pages = {337-355},
       adsurl = {https://ui.adsabs.harvard.edu/abs/1973A&A....24..337S}
}

@ARTICLE{Robitaille2006,
       author = {{Robitaille}, Thomas P. and {Whitney}, Barbara A. and {Indebetouw}, Remy and {Wood}, Kenneth and {Denzmore}, Pia},
        title = "{Interpreting Spectral Energy Distributions from Young Stellar Objects. I. A Grid of 200,000 YSO Model SEDs}",
      journal = {\apjs},
         year = 2006,
        month = dec,
       volume = {167},
       number = {2},
        pages = {256-285},
          doi = {10.1086/508424},
archivePrefix = {arXiv},
       eprint = {astro-ph/0608234},
 primaryClass = {astro-ph},
       adsurl = {https://ui.adsabs.harvard.edu/abs/2006ApJS..167..256R}
}

@ARTICLE{McClure2015,
       author = {{McClure}, M.~K. and {Espaillat}, C. and {Calvet}, N. and {Bergin}, E. and {D'Alessio}, P. and {Watson}, D.~M. and {Manoj}, P. and {Sargent}, B. and {Cleeves}, L.~I.},
        title = "{Detections of Trans-Neptunian Ice in Protoplanetary Disks}",
      journal = {\apj},
         year = 2015,
        month = feb,
       volume = {799},
       number = {2},
          eid = {162},
        pages = {162},
          doi = {10.1088/0004-637X/799/2/162},
archivePrefix = {arXiv},
       eprint = {1411.7618},
 primaryClass = {astro-ph.EP},
       adsurl = {https://ui.adsabs.harvard.edu/abs/2015ApJ...799..162M}
}

@ARTICLE{Loeffler2016,
       author = {{Loeffler}, M.~J. and {Moore}, M.~H. and {Gerakines}, P.~A.},
        title = "{The Effects of Experimental Conditions on the Refractive Index and Density of Low-temperature Ices: Solid Carbon Dioxide}",
      journal = {\apj},
         year = 2016,
        month = aug,
       volume = {827},
       number = {2},
          eid = {98},
        pages = {98},
          doi = {10.3847/0004-637X/827/2/98},
       adsurl = {https://ui.adsabs.harvard.edu/abs/2016ApJ...827...98L}
}

@ARTICLE{Poteet2011,
       author = {{Poteet}, Charles A. and {Megeath}, S. Thomas and {Watson}, Dan M. and {Calvet}, Nuria and {Remming}, Ian S. and {McClure}, Melissa K. and {Sargent}, Benjamin A. and {Fischer}, William J. and {Furlan}, Elise and {Allen}, Lori E. and {Bjorkman}, Jon E. and {Hartmann}, Lee and {Muzerolle}, James and {Tobin}, John J. and {Ali}, Babar},
        title = "{A Spitzer Infrared Spectrograph Detection of Crystalline Silicates in a Protostellar Envelope}",
      journal = {\apjl},
         year = 2011,
        month = jun,
       volume = {733},
       number = {2},
          eid = {L32},
        pages = {L32},
          doi = {10.1088/2041-8205/733/2/L32},
archivePrefix = {arXiv},
       eprint = {1104.4498},
 primaryClass = {astro-ph.SR},
       adsurl = {https://ui.adsabs.harvard.edu/abs/2011ApJ...733L..32P}
}

@ARTICLE{Crapsi2008,
       author = {{Crapsi}, A. and {van Dishoeck}, E.~F. and {Hogerheijde}, M.~R. and {Pontoppidan}, K.~M. and {Dullemond}, C.~P.},
        title = "{Characterizing the nature of embedded young stellar objects through silicate, ice and millimeter observations}",
      journal = {\aap},
         year = 2008,
        month = jul,
       volume = {486},
       number = {1},
        pages = {245-254},
          doi = {10.1051/0004-6361:20078589},
archivePrefix = {arXiv},
       eprint = {0801.4139},
 primaryClass = {astro-ph},
       adsurl = {https://ui.adsabs.harvard.edu/abs/2008A&A...486..245C}
}

@ARTICLE{Furlan2008,
       author = {{Furlan}, E. and {McClure}, M. and {Calvet}, N. and {Hartmann}, L. and {D'Alessio}, P. and {Forrest}, W.~J. and {Watson}, D.~M. and {Uchida}, K.~I. and {Sargent}, B. and {Green}, J.~D. and {Herter}, T.~L.},
        title = "{Spitzer IRS Spectra and Envelope Models of Class I Protostars in Taurus}",
      journal = {\apjs},
         year = 2008,
        month = may,
       volume = {176},
       number = {1},
        pages = {184-215},
          doi = {10.1086/527301},
archivePrefix = {arXiv},
       eprint = {0711.4038},
 primaryClass = {astro-ph},
       adsurl = {https://ui.adsabs.harvard.edu/abs/2008ApJS..176..184F}
}

@ARTICLE{Cordiner2025,
       author = {{Cordiner}, Martin A. and {Roth}, Nathan X. and {Kelley}, Michael S.~P. and {Bodewits}, Dennis and {Charnley}, Steven B. and {Drozdovskaya}, Maria N. and {Farnocchia}, Davide and {Micheli}, Marco and {Milam}, Stefanie N. and {Opitom}, Cyrielle and {Schwamb}, Megan E. and {Thomas}, Cristina A. and {Bagnulo}, Stefano},
        title = "{JWST Detection of a Carbon-dioxide-dominated Gas Coma Surrounding Interstellar Object 3I/ATLAS}",
      journal = {\apjl},
         year = 2025,
        month = oct,
       volume = {991},
       number = {2},
          eid = {L43},
        pages = {L43},
          doi = {10.3847/2041-8213/ae0647},
archivePrefix = {arXiv},
       eprint = {2508.18209},
 primaryClass = {astro-ph.EP},
       adsurl = {https://ui.adsabs.harvard.edu/abs/2025ApJ...991L..43C}
}

@ARTICLE{Gerakines1999,
       author = {{Gerakines}, P.~A. and {Whittet}, D.~C.~B. and {Ehrenfreund}, P. and {Boogert}, A.~C.~A. and {Tielens}, A.~G.~G.~M. and {Schutte}, W.~A. and {Chiar}, J.~E. and {van Dishoeck}, E.~F. and {Prusti}, T. and {Helmich}, F.~P. and {de Graauw}, Th.},
        title = "{Observations of Solid Carbon Dioxide in Molecular Clouds with the Infrared Space Observatory}",
      journal = {\apj},
         year = 1999,
        month = sep,
       volume = {522},
       number = {1},
        pages = {357-377},
          doi = {10.1086/307611},
       adsurl = {https://ui.adsabs.harvard.edu/abs/1999ApJ...522..357G}
}

@ARTICLE{EvD2025,
       author = {{van Dishoeck}, E.~F. and {Tychoniec}, {\L}. and {Rocha}, W.~R.~M. and {Slavicinska}, K. and {Francis}, L. and {van Gelder}, M.~L. and {Ray}, T.~P. and {Beuther}, H. and {Caratti o Garatti}, A. and {Brunken}, N.~G.~C. and {Chen}, Y. and {Devaraj}, R. and {Geers}, V.~C. and {Gieser}, C. and {Greene}, T.~P. and {Justtanont}, K. and {Le Gouellec}, V.~J.~M. and {Kavanagh}, P.~J. and {Klaassen}, P.~D. and {Janssen}, A.~G.~M. and {Navarro}, M.~G. and {Nazari}, P. and {Notsu}, S. and {Perotti}, G. and {Ressler}, M.~E. and {Reyes}, S.~D. and {Sellek}, A.~D. and {Tabone}, B. and {Tap}, C. and {Theijssen}, N.~C.~M.~A. and {Colina}, L. and {G{\"u}del}, M. and {Henning}, Th. and {Lagage}, P.-O. and {{\"O}stlin}, G. and {Vandenbussche}, B. and {Wright}, G.~S.},
        title = "{JWST Observations of Young protoStars (JOYS): Overview of program and early results}",
      journal = {\aap},
         year = 2025,
        month = jul,
       volume = {699},
          eid = {A361},
        pages = {A361},
          doi = {10.1051/0004-6361/202554444},
archivePrefix = {arXiv},
       eprint = {2505.08002},
 primaryClass = {astro-ph.GA},
       adsurl = {https://ui.adsabs.harvard.edu/abs/2025A&A...699A.361V}
}

@ARTICLE{Schutte1999,
       author = {{Schutte}, W.~A. and {Boogert}, A.~C.~A. and {Tielens}, A.~G.~G.~M. and {Whittet}, D.~C.~B. and {Gerakines}, P.~A. and {Chiar}, J.~E. and {Ehrenfreund}, P. and {Greenberg}, J.~M. and {van Dishoeck}, E.~F. and {de Graauw}, Th.},
        title = "{Weak ice absorption features at 7.24 and 7.41 MU M in the spectrum of the obscured young stellar object W 33A}",
      journal = {\aap},
         year = 1999,
        month = mar,
       volume = {343},
        pages = {966-976},
       adsurl = {https://ui.adsabs.harvard.edu/abs/1999A&A...343..966S}
}

@ARTICLE{Oba2010,
       author = {{Oba}, Yasuhiro and {Watanabe}, Naoki and {Kouchi}, Akira and {Hama}, Tetsuya and {Pirronello}, Valerio},
        title = "{Experimental Study of CO$_{2}$ Formation by Surface Reactions of Non-energetic OH Radicals with CO Molecules}",
      journal = {\apjl},
         year = 2010,
        month = apr,
       volume = {712},
       number = {2},
        pages = {L174-L178},
          doi = {10.1088/2041-8205/712/2/L174},
       adsurl = {https://ui.adsabs.harvard.edu/abs/2010ApJ...712L.174O}
}

@ARTICLE{Visser2009,
       author = {{Visser}, R. and {van Dishoeck}, E.~F. and {Doty}, S.~D. and {Dullemond}, C.~P.},
        title = "{The chemical history of molecules in circumstellar disks. I. Ices}",
      journal = {\aap},
         year = 2009,
        month = mar,
       volume = {495},
       number = {3},
        pages = {881-897},
          doi = {10.1051/0004-6361/200810846},
archivePrefix = {arXiv},
       eprint = {0901.1313},
 primaryClass = {astro-ph.SR},
       adsurl = {https://ui.adsabs.harvard.edu/abs/2009A&A...495..881V}
}

@ARTICLE{Cook2011,
       author = {{Cook}, A.~M. and {Whittet}, D.~C.~B. and {Shenoy}, S.~S. and {Gerakines}, P.~A. and {White}, D.~W. and {Chiar}, J.~E.},
        title = "{The Thermal Evolution of Ices in the Environments of Newly Formed Stars: The CO$_{2}$ Diagnostic}",
      journal = {\apj},
         year = 2011,
        month = apr,
       volume = {730},
       number = {2},
          eid = {124},
        pages = {124},
          doi = {10.1088/0004-637X/730/2/124},
       adsurl = {https://ui.adsabs.harvard.edu/abs/2011ApJ...730..124C}
}

@ARTICLE{Bergner2024,
       author = {{Bergner}, Jennifer B. and {Sturm}, J.~A. and {Piacentino}, Elettra L. and {McClure}, M.~K. and {{\"O}berg}, Karin I. and {Boogert}, A.~C.~A. and {Dartois}, E. and {Drozdovskaya}, M.~N. and {Fraser}, H.~J. and {Harsono}, Daniel and {Ioppolo}, Sergio and {Law}, Charles J. and {Lis}, Dariusz C. and {McGuire}, Brett A. and {Melnick}, Gary J. and {Noble}, Jennifer A. and {Palumbo}, M.~E. and {Pendleton}, Yvonne J. and {Perotti}, Giulia and {Qasim}, Danna and {Rocha}, W.~R.~M. and {van Dishoeck}, E.~F.},
        title = "{JWST Ice Band Profiles Reveal Mixed Ice Compositions in the HH 48 NE Disk}",
      journal = {\apj},
         year = 2024,
        month = nov,
       volume = {975},
       number = {2},
          eid = {166},
        pages = {166},
          doi = {10.3847/1538-4357/ad79fc},
archivePrefix = {arXiv},
       eprint = {2409.08117},
 primaryClass = {astro-ph.EP},
       adsurl = {https://ui.adsabs.harvard.edu/abs/2024ApJ...975..166B}
}

@ARTICLE{Dartois2022,
       author = {{Dartois}, E. and {Noble}, J.~A. and {Ysard}, N. and {Demyk}, K. and {Chabot}, M.},
        title = "{Influence of grain growth on CO$_{2}$ ice spectroscopic profiles. Modelling for dense cores and disks}",
      journal = {\aap},
         year = 2022,
        month = oct,
       volume = {666},
          eid = {A153},
        pages = {A153},
          doi = {10.1051/0004-6361/202243929},
archivePrefix = {arXiv},
       eprint = {2207.09411},
 primaryClass = {astro-ph.GA},
       adsurl = {https://ui.adsabs.harvard.edu/abs/2022A&A...666A.153D}
}

@ARTICLE{Kemper2004,
       author = {{Kemper}, F. and {Vriend}, W.~J. and {Tielens}, A.~G.~G.~M.},
        title = "{The Absence of Crystalline Silicates in the Diffuse Interstellar Medium}",
      journal = {\apj},
         year = 2004,
        month = jul,
       volume = {609},
       number = {2},
        pages = {826-837},
          doi = {10.1086/421339},
archivePrefix = {arXiv},
       eprint = {astro-ph/0403609},
 primaryClass = {astro-ph},
       adsurl = {https://ui.adsabs.harvard.edu/abs/2004ApJ...609..826K}
}

@ARTICLE{Hollenbach2009,
       author = {{Hollenbach}, David and {Kaufman}, Michael J. and {Bergin}, Edwin A. and {Melnick}, Gary J.},
        title = "{Water, O$_{2}$, and Ice in Molecular Clouds}",
      journal = {\apj},
         year = 2009,
        month = jan,
       volume = {690},
       number = {2},
        pages = {1497-1521},
          doi = {10.1088/0004-637X/690/2/1497},
archivePrefix = {arXiv},
       eprint = {0809.1642},
 primaryClass = {astro-ph},
       adsurl = {https://ui.adsabs.harvard.edu/abs/2009ApJ...690.1497H}
}

@ARTICLE{Oberg2011b,
       author = {{{\"O}berg}, Karin I. and {Boogert}, A.~C. Adwin and {Pontoppidan}, Klaus M. and {van den Broek}, Saskia and {van Dishoeck}, Ewine F. and {Bottinelli}, Sandrine and {Blake}, Geoffrey A. and {Evans}, II, Neal J.},
        title = "{The Spitzer Ice Legacy: Ice Evolution from Cores to Protostars}",
      journal = {\apj},
         year = 2011,
        month = oct,
       volume = {740},
       number = {2},
          eid = {109},
        pages = {109},
          doi = {10.1088/0004-637X/740/2/109},
archivePrefix = {arXiv},
       eprint = {1107.5825},
 primaryClass = {astro-ph.GA},
       adsurl = {https://ui.adsabs.harvard.edu/abs/2011ApJ...740..109O}
}

@ARTICLE{Gross2025,
       author = {{Gross}, Rachel E. and {Yang}, Yao-Lun and {Cleeves}, L. Ilsedore and {van Dishoeck}, Ewine F. and {Garrod}, Robin T. and {Jin}, Mihwa and {Sakai}, Nami and {Shingledecker}, Christopher N. and {Kim}, JaeYeong and {Bergner}, Jennifer B. and {Evans}, II, Neal J. and {Green}, Joel D. and {Kim}, Chul-Hwan and {Lee}, Jeong-Eun and {Okoda}, Yuki and {Rocha}, Will R.~M. and {Shope}, Brielle and {Tyagi}, Himanshu},
        title = "{CORINOS IV: Quantifying Baseline-Fitting Uncertainties in SO$_2$ Ice Measurements with JWST/MIRI}",
      journal = {arXiv e-prints},
         year = 2025,
        month = dec,
          eid = {arXiv:2512.20820},
        pages = {arXiv:2512.20820},
          doi = {10.48550/arXiv.2512.20820},
archivePrefix = {arXiv},
       eprint = {2512.20820},
 primaryClass = {astro-ph.EP},
       adsurl = {https://ui.adsabs.harvard.edu/abs/2025arXiv251220820G}
}

@misc{Kim2025,
       author = {{Kim}, J.},
  title  = {{in prep.}},
  year   = {2025},
  note   = {in prep.}
}
\bibliographystyle{aasjournal}

\end{document}